\documentclass[aip,jcp]{revtex4-1}
\usepackage{graphicx}
\usepackage{subfigure}
\usepackage{caption}
\usepackage{multirow}
\usepackage{docs}%
\usepackage{dcolumn}
\usepackage{bm}
\usepackage{algorithm}
\usepackage{algpseudocode}
\usepackage{epstopdf}
\usepackage{esvect}
\usepackage[colorlinks=true,linkcolor=blue]{hyperref}
\usepackage[titletoc,title]{appendix}
\draft 
\expandafter\ifx\csname package@font\endcsname\relax\else
 \expandafter\expandafter
 \expandafter\usepackage
 \expandafter\expandafter
 \expandafter{\csname package@font\endcsname}%
\fi

\begin{document}

\title{A transferable artificial neural network model for atomic forces in nanoparticles}
\author{Shweta Jindal}
\author{Satya S. Bulusu}
\email[]{sbulusu@iiti.ac.in}
\affiliation{$^1$Discipline of Chemistry, Indian Institute of Technology Indore, Simrol, Indore 453552, India}%
\begin{abstract} 
We have designed a new method to fit the energy and atomic forces using a single artificial neural network (SANN) for any number of chemical species present in a molecular system. The traditional approach for fitting the potential energy surface (PES) for a multicomponent (MC) system using artificial neural network (ANN) is to consider $n$ number of networks for $n$ number of chemical species in the system. This shoots the computational cost and makes it difficult to apply to a system containing more number of species. We present a new strategy of using a SANN to compute energy and forces of a chemical system. Since, atomic forces are significant for geometry optimizations and molecular dynamics simulations (MDS) for any chemical system, their accurate prediction is of utmost importance. So, to predict the atomic forces, we have modified the traditional way of fitting forces from underlying energy expression. We have applied our strategy to study geometry optimizations and dynamics in gold-silver nanoalloys and thiol protected gold nanoclusters. Also, force fitting has made it possible to train smaller size systems and extrapolate the parameters  to make accurate predictions for larger systems. This proposed strategy has definitely made the mapping and fitting of atomic forces easier and can be applied to a wide variety of molecular systems. 
\end{abstract}
\pacs{}

\maketitle 
\section{INTRODUCTION}
In order to reduce the computational complexities of quantum mechanical (QM) calculations, data learning techniques have been widely applied to get the accuracy of QM calculations in a time of few seconds. Machine learning\cite{deepNN,MLT_1,gdml,behlertut,use_ann1,use_ann2,sphharmjcp,markus_multi,use_ann3,powerspectrum,rupp,botu,snap_pot,mybispec} has specially garnered attention due to its data dependent parameter fitting and extremely low errors in predicting the desired property of interest. Among the machine learning techniques, ANN has been practised by various research groups\cite{behlertut,use_ann1,use_ann2,sphharmjcp,markus_multi,use_ann3} as a robust technique to fit the potential energy surface (PES) of a molecular system. In order to fit a property of a molecular system, the environment of an atom is converted into numerical parameters called as \textit{descriptors}.\cite{behlertut,powerspectrum,sphharmjcp,rupp,botu,snap_pot,mybispec} For molecular systems containing a single component (SC), a lot of techniques\cite{behlertut,sphharmjcp,powerspectrum} have been introduced which have been efficiently implemented by using descriptors and have mimicked the chemical systems with high accuracy. On the other hand, representing MC system by descriptors is a challenging problem, which has been taken up by several research groups\cite{nong_multi,weighted_Acsf,markus_multi} recently. Many of them\cite{nong_multi,weighted_Acsf,markus_multi} have fitted only the energy for a system and have not fitted atomic forces. Behler et al.,\cite{behler_mc1,behler_mc2} introduced the fitting of energy and atomic forces for MC system via ANN, by using $n$ networks for $n$ number of chemical species present in the system. This makes it difficult to use for systems containing a lot of different chemical species as the computational cost for fitting increases with increase in $n$. In a MC system capturing of atomic behaviour via descriptors is necessary as atomic forces play an important role\citep{vecforce_GPR,ontheforce,internalforce,unary_binary_F,botuforce_prb,botuforce_jpcc} in optimizations, MDS and other applications requiring the calculation of gradients of energy. 

In the present work, we have achieved a conjoint but a decoupled fitting of energy and forces of a MC system using a SANN for all the elements in a system. We used optimal number of neurons and input functions for describing an atomic environment. We have used higher order invariants (HOI)\cite{sphharmjcp}, called power spectrum coefficients as the descriptors. The proposed method can be easily extrapolated to another HOI based descriptors which are bispectrum coefficients\cite{mybispec,snap_pot} which is currently in progress. The forces are not calculated directly from the energy obtained for a system. Instead they are calculated using a dummy energy and therefore we term our fitting of energy and forces as concurrent but decoupled. This fitting is computationally cheap as we use just one network for all the types of elements in a system. In this way, we can attain massive speed ups to run the MDS of a MC system. In addition, this strategy helps to train atomic forces for smaller atomic systems and use them to predict precise atomic forces for identical environments in larger atomic systems. Since the introduction of descriptor-ANN model by Behler et al.,\cite{behler_prl} we have come a long way to generalise the technique for metallic nanoclusters\cite{siva_sodium,siva_gold,sphharmjcp,agau_jcp} to any type of nanoparticles now. The key points of our proposed strategy are (a) Due to a precise atomic force fitting, the ANN weights fitted for a small size system can be extrapolated to similar compositions in the larger size systems, (b) The overall model is transferable in two ways - (i) It can be used to fit any nanoparticle's PES as we require just a dataset, from which, the inter-atomic distances, effective nuclear charges and reduced mass can be utilised to give an input to ANN, (ii) The weights are transferable for a chemical system to any size of the clusters of similar composition.

Since the complexity of a descriptor increases with the type of chemical species, therefore, we tested our proposed technique for two different systems - bimetallic nanoalloys made up of silver and gold atoms  $(AgAu)_{55}$-$(AgAu)_{147}$, and thiol protected gold nanoclusters $Au_{13}(SH)_{6}$-$Au_{38}(SH)_{24}$. We have done geometry optimizations and MDS for studying the dynamics of $Ag_{35}Au_{112}$ and $Au_{68}(SH)_{32}$. We got some important insights about structural stability of these clusters. We also found out that $Ag_{35}Au_{112}$ diverges from icosahedron geometry and adopts an amorphous structure which is identical to amorphous global minimum of $Au_{147}$.\cite{sphharmjcp} 

We have discussed the existing optimization approach for SC system using ANN in Section \ref{sc_sys}, the proposed method for decoupled and concurrent fitting of energy and forces for MC system in Section \ref{mc_sys}. A general algorithm to generate a dataset is briefly discussed in Section \ref{gendata} followed by the computational details for the systems fitted in Section \ref{compu_det}, results and discussion in Section \ref{result} and conclusions in Section \ref{conclu}.

\section{THEORY}
\subsection{Existing optimization approach for SC systems using HOI descriptors}\label{sc_sys}
The local environment of an atom $h$, can be well represented as a summation of delta functions,\cite{taylor,powerspectrum} which is unity at any neighbouring atom $k$ and zero elsewhere. In order to define the local environment of an atom, we use a cut-off function ($f_c(r_{hk})$) which allows us to consider only a finite number of neighbouring atoms.
\begin{equation}\label{delta density}
\rho(\textbf{r}) = \sum_{h\neq k}e^{-nr_{hk}^2}\delta (\textbf{r}-\textbf{r}_{hk})f_c(r_{hk})
\end{equation}
where, $\rho(\textbf{r})$ denotes the atomic density function (ADF) for an atom. In order to consider the impact of inter atomic distance on the parameters such as energy, obtained through the above ADF, we consider exponential weighting\cite{sphharmjcp,taylor} with parameter $n$. This ensures that as the inter atomic distance increases, the impact of the bonding decreases. The cut-off function\cite{behlertut} $f_c(r_{hk})$ is given as
\begin{equation}\label{cutoff}
f_c(r_{hk}) = \frac{1}{2}\left[\cos\left(\frac{\pi r_{hk}}{r_c}\right)+1\right] 
\end{equation} 
where, $\textbf{r}_{hk}$ is a vector defined as $\textbf{r}_h - \textbf{r}_k$, ${r}_{hk}$ is the magnitude of $\textbf{r}_{hk}$ and $r_c$ is the cut off radius. We use a cut off radius of $8$\AA , which we have decided after comparing the error in the predicted energy for several choices of cut off radius. 

We note that spherical harmonics ($Y_{lm}$) is an orthonormal basis set for $L_2$ (square integrable) functions on the unit sphere\cite{powerspectrum}. Hence, we can expand ADF as follows.
\begin{equation} \label{density expand in SH}
\rho(\hat{\textbf{r}}) = \sum_{l=0}^{\infty}\sum_{m=-l}^{l}c_{nlm}Y_{lm}(\hat{\textbf{r}_{hk}})
\end{equation}
where, $\hat{\textbf{r}_{hk}}$ is the unit vector of $\textbf{r}_{hk}$. $c_{nlm}$ in the above equation can be obtained as inner product of $\rho(\hat{\textbf{r}})$ and $Y_{lm}$, which after basic algebraic manipulations is given as
\begin{equation}
c_{nlm}^h = \sum_{h\neq k} Y_{lm}^{*}(\hat{\textbf{r}_{hk}})e^{-nr_{hk}^2}f_c(r_{hk})
\end{equation}
We observe that the spherical harmonics coefficients $c_{nlm}$ contain the complete information of the ADF. At a particular frequency \textit{l} and degree \textit{m}, $c_{nlm}$ can help obtain the amplitude and phase of the function. We can obtain a rotationally, permutationally and reflection invariant descriptor, referred to as power spectrum\cite{powerspectrum,sphharmjcp}, as follows. 
\begin{equation}\label{power spectrum}
P_{nl}^h = \frac{4\pi}{2l+1}\sum_{m=-l}^l c_{nlm}^{h*}c_{nlm}^h
\end{equation}
$P_{nl}^h$ is called as the power spectrum for an atom $h$, since it specifies the content of the projected function which is ADF in our case at a particular frequency $l$. Traditionally, the power spectrum is obtained as the fourier transform of the autocorrelation function. 

Although, power spectrum contains the entire information of the local atomic environment, in order to provide exact radial environment around an atom, we also take radial functions as descriptors given in equation \ref{radialfunctio}. The falling of the gaussian function is controlled by the parameter $\xi$ and the neighbours are limited by cut off function $f_c(r_{hk})$.
\begin{equation}\label{radialfunctio}
d_{rad}^{h} = \sum_{h\neq k} e^{-\xi r_{hk}^2}f_c(r_{hk})
\end{equation}
Now, in order to obtain the energy of a cluster from descriptor functions we have applied ANN as the kernel.\cite{powerspectrum,rupp,botu,behlertut} Subsequently, the atomic forces can be obtained as a derivative of the energy of the cluster. For a two hidden layer ($i$ and $j$) network, the energy of an atom in terms of neural network weights ($\omega_{ki}^{01}$, $\omega_{ij}^{12}$, $\omega_{j1}^{23}$, $\tau_i$, $\tau_j$) and input descriptors ($d_{N,k}$) is given as
\begin{equation}\label{NNEatom}
E_{N}= \sum_{j=1}^{N_{hl_{2}}} \omega_{j1}^{23}.f_j \left(  \tau_j+\sum_{i=1}^{N_{hl_{1}}} \omega_{ij}^{12}.f_i \left( \tau_i+\sum_{k=1}^{N_{input}} \omega_{ki}^{01}.d_{N,k} \right)  \right)
\end{equation}
where, $f_i$ and $f_j$ are the sigmoid functions for hidden layer $i$ and $j$, $\omega_{ki}^{01}$, $\omega_{ij}^{12}$, $\omega_{j1}^{23}$ are the weights from input layer to first hidden layer $i$, first hidden layer $i$ to second hidden layer $j$ and second hidden layer $j$ to output layer, respectively. $\tau_i$ and $\tau_j$ are the bias weights for $i$ and $j$ layers, respectively. $N_{hl_{1}}$ and $N_{hl_{2}}$ are the number of neurons in layer $i$ and $j$, respectively and $N_{input}$ are the number of input descriptor functions for an atom.

The total energy of a cluster is then calculated as the sum of atomic energies, $E_{cluster} = \sum_{N=1}^{atoms} E_{N}$. Since atomic forces are a vector quantity, it is given as negative gradient of energy with respect to each component of the 
$\textbf{r}_{hk}$ vector. As energy calculation is dependent on descriptors which are functions of atomic positions, therefore, using chain rule, forces are calculated as follows, where, $\alpha \in \{x,y,z\}$.
\begin{equation}\label{forceeq}
F_{\alpha} =-\frac{\partial E_{cluster}}{\partial {\alpha}}
 =-\sum_{N=1}^{atoms} \frac{\partial E_N}{\partial {\alpha}} \\
 =-\sum_{N=1}^{atoms} \sum_{k=1}^{input} \frac{\partial E_N}{\partial d_{N,k}}\frac{\partial d_{N,k}}{\partial {\alpha}}
\end{equation}
We have used global extended Kalman filter\cite{symfunjcp,witkoskie} to optimize the ANN weights whose implementation details can be found in our earlier work.\cite{symfunjcp} We chose Kalman filter as the optimization algorithm as its a highly robust technique for fitting a dataset consisting of a high numerical variations. Since, our aim is to concurrently fit the energy and forces, the error vector $f_{err}^{cluster}$ is of dimension $(3N+1) \times 1$ (N is the total number of atoms in a cluster) which is given as 
\begin{equation}\label{errorvec}
f_{err}^{cluster} =\left[ E_{DFT}-E_{cluster}^{ANN},F_{\alpha_{1}}^{DFT}-F_{\alpha_{1}}^{ANN},..,F_{\alpha_{3N}}^{DFT}-F_{\alpha_{3N}}^{ANN}\right]
\end{equation} 
where, $E_{DFT}$ and $E_{cluster}^{ANN}$ are the energy of a cluster obtained from DFT and ANN, respectively. $F_{\alpha_{N}}^{DFT}$ and $F_{\alpha_{N}}^{ANN}$ are the atomic forces at a particular coordinate obtained from DFT and ANN, respectively.
\subsection{A new approach for MC systems using HOI descriptors}\label{mc_sys}
The modelling of ADF for a MC system cannot be the same as SC system because each element in the periodic table has different bonding patterns which is not captured by kernel based methods. Behler et al.\cite{behler_mc2,behler_mc1} proposed to use different set of network weights for all the elemental species in a system. This scheme is computationally effective for a system consisting of less number of chemical species. On increasing the types of chemical species, the number of networks increases, thus increasing the complexity of fitting. One way to overcome this situation is to differentiate the chemical species at the descriptor level and use a single set of network weights for the entire molecular system. Recently, Gastegger et al.,\cite{weighted_Acsf} Artrith et al.\cite{nong_multi} and Unke et al.\cite{markus_multi} have proposed the weighting of descriptor functions according to an element and fitted the energies of a molecular dataset.

Since atomic forces are of utmost importance to run MDS, we aim to concurrently fit energy and forces for MC system using ANN which has currently not been done using a single network to the best of our knowledge. For this, we first propose a bond specific weighting of ADF and radial functions (given in equation \ref{delta density} and \ref{radialfunctio}) as 
\begin{equation}\label{bond weighted delta density}
\rho_{mod}^h(\textbf{r}) = \sum_{h\neq k}w_{hk} e^{-nr_{hk}^2}\delta (\textbf{r}-\textbf{r}_{hk})f_c(r_{hk})
\end{equation}
\begin{equation}\label{weighted radialfunctio}
d_{\text{mod rad}}^{h} = \sum_{h\neq k}w_{hk} e^{-\xi r_{hk}^2}f_c(r_{hk})
\end{equation}
The $w_{hk}$ is specific for a bond ($b_{hk}$) between atom $h$ and $k$. This is chosen as $exp({\frac{\mu_{hk}}{m_\alpha}})$, where $\mu_{hk}$ is the reduced mass of $b_{hk}$ given as $\frac{m_h\times m_k}{m_h+m_k}$, $m_h$, $m_k$ are the molecular mass of atom $h$ and $k$, respectively. $m_\alpha$ takes the value of the molecular mass of atom whose local environment is being calculated. Comparing with the existing approach for SC systems, it can be deduced as modified exponentially weighted descriptors  (MEAD). We calculate energy of the cluster using MEAD in the ANN. 

Next, to obtain the atomic forces for the MC system, we have to use the gradient of energy which is obtained from MEAD above. This leads to an unnecessary scaling of the forces as the individual weighting of bonds makes it difficult for ANN to find a global minimum in weights. We propose to alleviate this bottleneck by modelling the forces in a decoupled manner from the energy obtained above. For this, we set the bond specific weighting in MEAD to be unity. This makes the MC system to behave as SC system. We calculate a dummy energy from the ANN using the descriptors with bond specific weighting as unity. These descriptors does not contain any element specific information. So, to incorporate the nature of the atom in the local environment, we propose an element specific weighting of the gradients of the descriptor with respect to the coordinates of the atoms as shown in equation \ref{modified force}. This element specific weighting embeds the fluctuations in the descriptor with slight variation in position with respect to a particular element such that when the data will be trained via ANN, network will recognise the element specific variations for forces.
\begin{equation}\label{modified force}
F_{\alpha}^{\text{weighted}}
 =-\sum_{N=1}^{atoms} \sum_{k=1}^{input} \frac{\partial E_N^{'}}{\partial d_{N,k}}\left(w_{\beta N}\times\frac{\partial d_{N,k}}{\partial {\alpha}}\right)
\end{equation}
Here, $E_N^{'}$ is the dummy energy obtained from descriptors with $w_{hk} = 1$ and $d_{N,k}$ are the MEAD with $w_{hk} = 1$. The $E_N^{'}$ is termed as a dummy energy as no element specific information is contained and we do not fit the energy of the cluster using this. $w_\beta$ is chosen as the ratio of the effective nuclear charge of the valence electrons of an element ($Z_{e}$) to that of total effective nuclear charge of all the chemical species present in the molecular system. 

\begin{figure}[htp]
\centering
\includegraphics[width=0.8\textwidth ,height=0.6\textwidth]{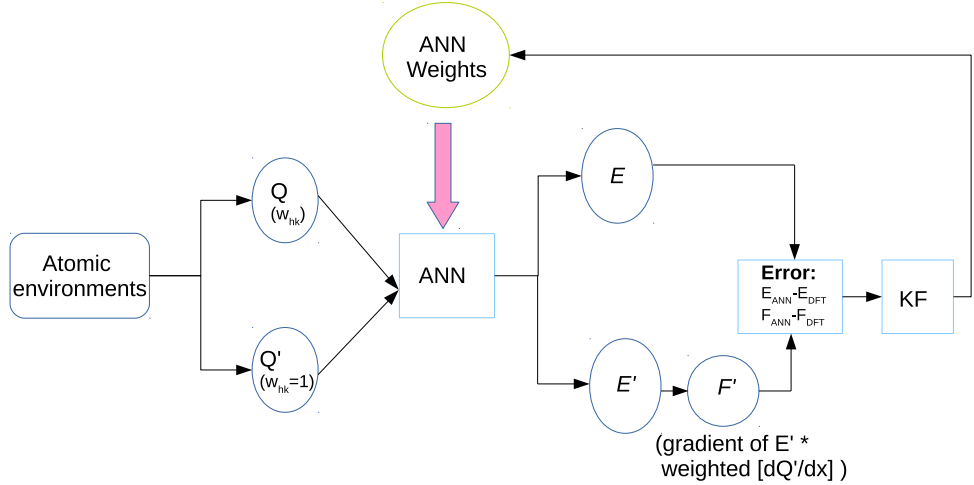}
\caption{The proposed model for concurrently fitting energy and forces of a cluster. Q and Q' are the descriptors for $E$ and $F'$, respectively. $F' = F_{\alpha}^{\text{weighted}}$ and $E = E_{cluster}$. The block KF represents the global extended Kalman filter.}\label{bispec_val_plot}
\end{figure}

Our proposed model is shown in Figure \ref{bispec_val_plot}. It consists of supplying two set of decriptors - (i) MEAD for energy, (ii) MEAD with $w_{hk} = 1$  for forces, to the ANN. Using the weights of the ANN and the descriptor for the energy we obtain the system energy as first output of the model. And, in parallel, using the same weights of the ANN and the descriptor for the forces we obtain the atomic forces. Global extended Kalman filter optimizes the weights of ANN to minimize the norm of the error vector (equation \ref{errorvec}). We have also summarised our model in an algorithm as shown in Algorithm \ref{multomalgo}. The descriptor calculating codes and the ANN codes were made in-house in fortran 90.

\begin{algorithm}[H]
\caption{Decoupled fitting of E and F for MC system}
\label{multomalgo}
\begin{algorithmic}[1]
\State Calculate the MEAD using Eq. (\ref{bond weighted delta density}) and (\ref{weighted radialfunctio}) of all the atoms for fitting the energy of a cluster.
\State Calculate the descriptors (MEAD with $w_{hk} = 1$) and gradients of descriptors with respect to atomic positions using $\left(w_\beta\times\frac{\partial d_{N,k}}{\partial {\alpha}}\right)$ for all the atoms for fitting the atomic forces.
\State Split the entire dataset into a training and testing set for ANN.
\State The ANN is then trained with different types of clusters and the weights are validated after each iteration.
\State The training is stopped when a minimum root mean square error is observed.
\end{algorithmic}
\end{algorithm}
\section{A general way to generate initial data for any atomic system}\label{gendata}
Since, our proposed method utilises the inter-atomic distances, effective nuclear charges of the chemical species and the reduced masses corresponding to a particular bond, we can use this to fit any atomic system's PES. The basic ingredient for any PES generation is the dataset, which is fitted using descriptor-ANN integrated model. A generalise step by step model to generate any dataset is given in Algorithm \ref{datagenalgo}.
\begin{algorithm}[H]
\caption{A step by step way for data generation}
\label{datagenalgo}
\begin{algorithmic}[1]
\State Choose an initial potential - Empirical potential, Force fields or \textit{ab initio}.
\State Run Monte Carlo simulations, Basin hopping optimizations, MDS using any of the above potentials.
\State Collect around 2000 clusters and optimize them for minimizing the forces.
\State Calculate the \textit{descriptors} and its gradients for the clusters and simultaneously calculate the energy and forces using DFT.
\State Feed the data obtained in above step to ANN and fit it using Kalman filter/Back propagation/Conjugate gradient/Lavenburg Marquardt/Quasi Newton BFGS or any other algorithm to optimise a set of weights.
\State Run MDS integrated with ANN weights at different temperatures to generate more dataset for an accurate representation of PES.
\State In case of big clusters (\textgreater 100 atoms), split them in small atomic environments such that core and surface configurations are included in the dataset.
\State On refining and optimizing the different clusters obtained, repeat step 4.
\State A final fitting is done using around 11000 clusters and the converged weights can be utilised further in many applications.
\end{algorithmic}
\end{algorithm}
\section{COMPUTATIONAL DETAILS}\label{compu_det}
In order to substantiate the proposed theory, we have fitted potential energy surface and the forces consistent with it for gold-silver nanoalloys and thiol protected gold nanoclusters, and have evaluated their dynamics. 
\subsection{Parameters for fitting $(AgAu)_{55}$ - $(AgAu)_{147}$}\label{fit_agau}
In order to generate training data for $(AgAu)_{55}$ - $(AgAu)_{147}$, an initial data consisting of $(AgAu)_{55}$ was generated using Gupta potential\cite{gupta1} as the inter atomic potential in molecular dynamics simulation (MDS). After getting around 2500 clusters, an initial run of training was performed. Using the obtained set of ANN weights as the inter-atomic potential, MDS was run at 300 K, 400 K, 500 K and 600 K at a time step of 1\textit{fs} for $(AgAu)_{55}$ and $(AgAu)_{147}$. To avoid high computational costs for generating ab initio data of $(AgAu)_{147}$, we split\citep{sphharmjcp} around 1000 clusters of $(AgAu)_{147}$ into different atomic environments. A total data of 11,000 clusters was accumulated containing different compositions of $(AgAu)_{55}$ and various environments of $Ag_{35}Au_{112}$. We have taken a composition of 24 \% of silver atoms in $(AgAu)_{147}$, as it promises to be catalytically dynamic.\cite{agau_jcp} The energy and forces calculations for the dataset was executed on Vienna Ab initio Simulation Package (VASP).\cite{vasp1,vasp2,vasp3,vasp4} Scalar relativistic effects and the core electrons are taken care of by projector augmented wave (PAW) method. Generalized gradient approximation and Perdew-Burke-Ernzerhof (PBE)\cite{pbe1,pbe2} functional is applied for treating electron correlations. Gamma k-point (1$\times$1$\times$1) mesh is used to sample the Brillouin zone. The threshold energy is set as 260 eV and the force convergence is set as $10^{-4}$. A box length of $22\times 22\times 22 \AA^{3}$ is applied for the entire dataset with a vacuum dimension of 11 \AA . The dataset was splitted in a training set of 9,500 clusters and a testing set of 1,500 clusters. The number of inputs defining the environment for an atom was 59 which was obtained by taking $l$ from $0$ to $9$ in eq. \ref{power spectrum}. The $n$ in eq. \ref{bond weighted delta density} takes on 5 values in order to make the function fall smoothly with increasing inter-atomic distance. In this work, we have fixed the $n$ values to be $0.0028, 0.0040, 0.0110, 0.0280$ and $0.059$. Further, we took 9 radial functions in eq. \ref{radialfunctio} by considering 9 values of $\xi$ to be $0.005, 0.015, 0.0230, 0.038, 0.060, 0.090, 0.150, 0.260$ and $0.480$. The number of hidden layer neurons were set to be 30 i.e., in eq. \ref{NNEatom}, $N_{hl_{1}}$ and $N_{hl_{2}}$ is 30 each. The $w_\beta$ value for $Au$ and $Ag$ in Eq. \ref{modified force} is calculated using Clementi - Raimondi\cite{effnuccharge1,effnuccharge2} effective nuclear charges.
\subsection{Parameters for fitting $Au_m(SH)_n$}
We have taken a diverse set of $Au_m(SH)_n$ clusters in which $m$ varies from $13$ to $38$ and $n$ varies from $6$ to $24$ to fit the energy and forces. Since $Au_{13}(SH)_{6}$, $Au_{13}(SH)_{8}$, $Au_{13}(SH)_{9}$ and $Au_{15}(SH)_{8}$ are small sized clusters, therefore, we generated the initial data containing these composition clusters by MDS coupled with density functional theory (DFT). After getting an initial data, we generated ANN weights and then integrated them with MDS for generating more data for rest of the compositions in span of $Au_{13}(SH)_{6}$ to $Au_{38}(SH)_{24}$. Overall, we generated 11,500 clusters and divided them into a training data set of 10,000 clusters and a testing data set of 1,500 clusters. The DFT calculations were performed on VASP. All the parameters related to DFT calculations were same as discussed in Section \ref{fit_agau}. The number of inputs per atom was kept 59 for all the $Au_m(SH)_n$ clusters. The $n$ and $\xi$ values were kept same as mentioned in Section \ref{fit_agau}. The network for $Au_m(SH)_n$ also had 30 neurons in both the hidden layers. The $w_\beta$ values for $Au$, $S$ and $H$ in Eq. \ref{modified force} is calculated using Clementi - Raimondi\cite{effnuccharge1,effnuccharge2} effective nuclear charges.
\section{RESULTS AND DISCUSSION}\label{result}
\subsection{Silver-Gold nanoalloys: Study of $Ag_{35}Au_{112}$}
\begin{figure}[htp]
\centering
\includegraphics[width=4.4in,height=2.3in]{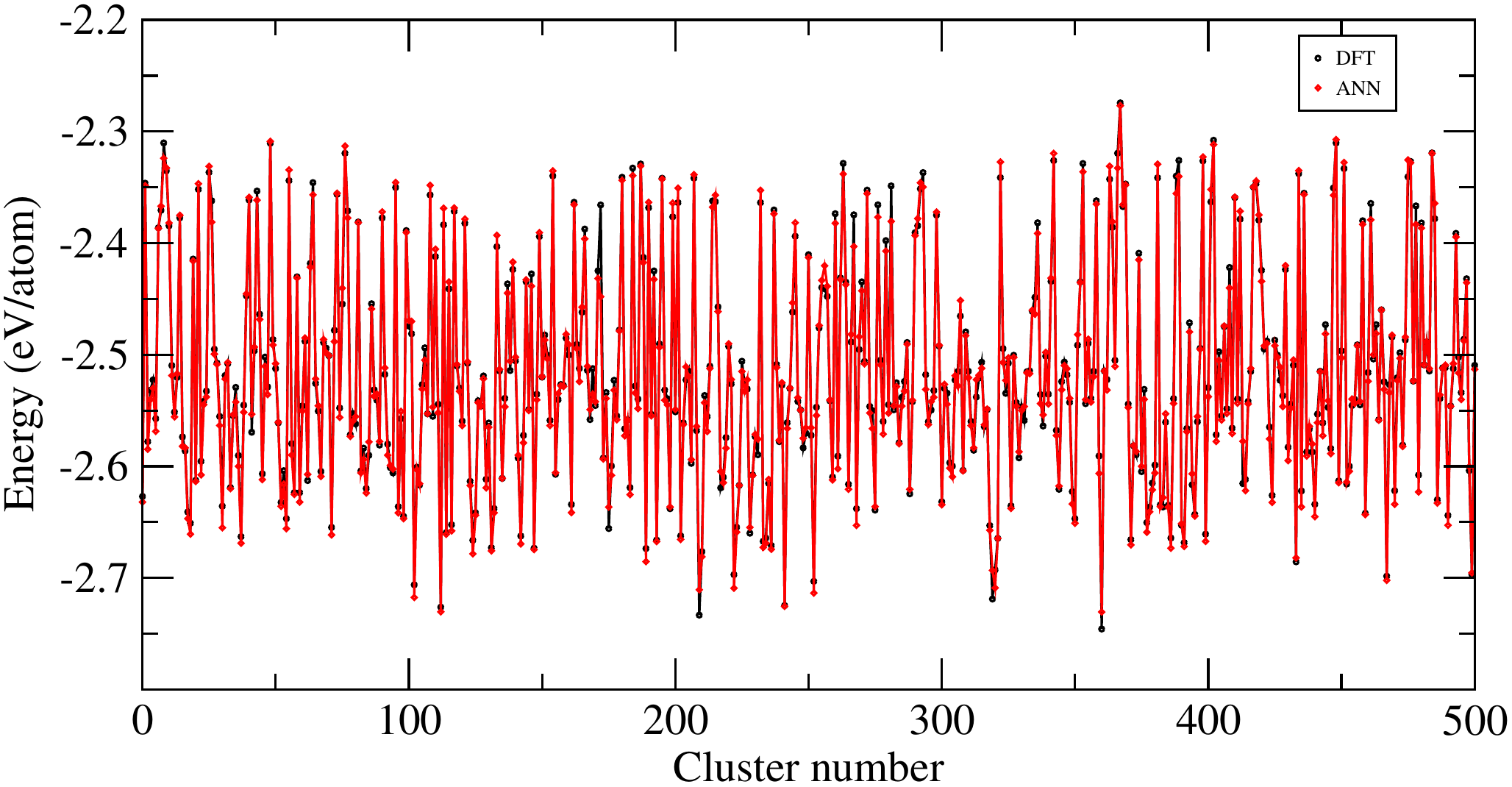}
\caption{Comparison of DFT and ANN predicted energies for $(AgAu)_{55}$ - $(AgAu)_{147}$}
\label{comp_dftann_agau}
\end{figure}

\begin{figure}[htp]
\centering
\includegraphics[width=4.4in,height=2.3in]{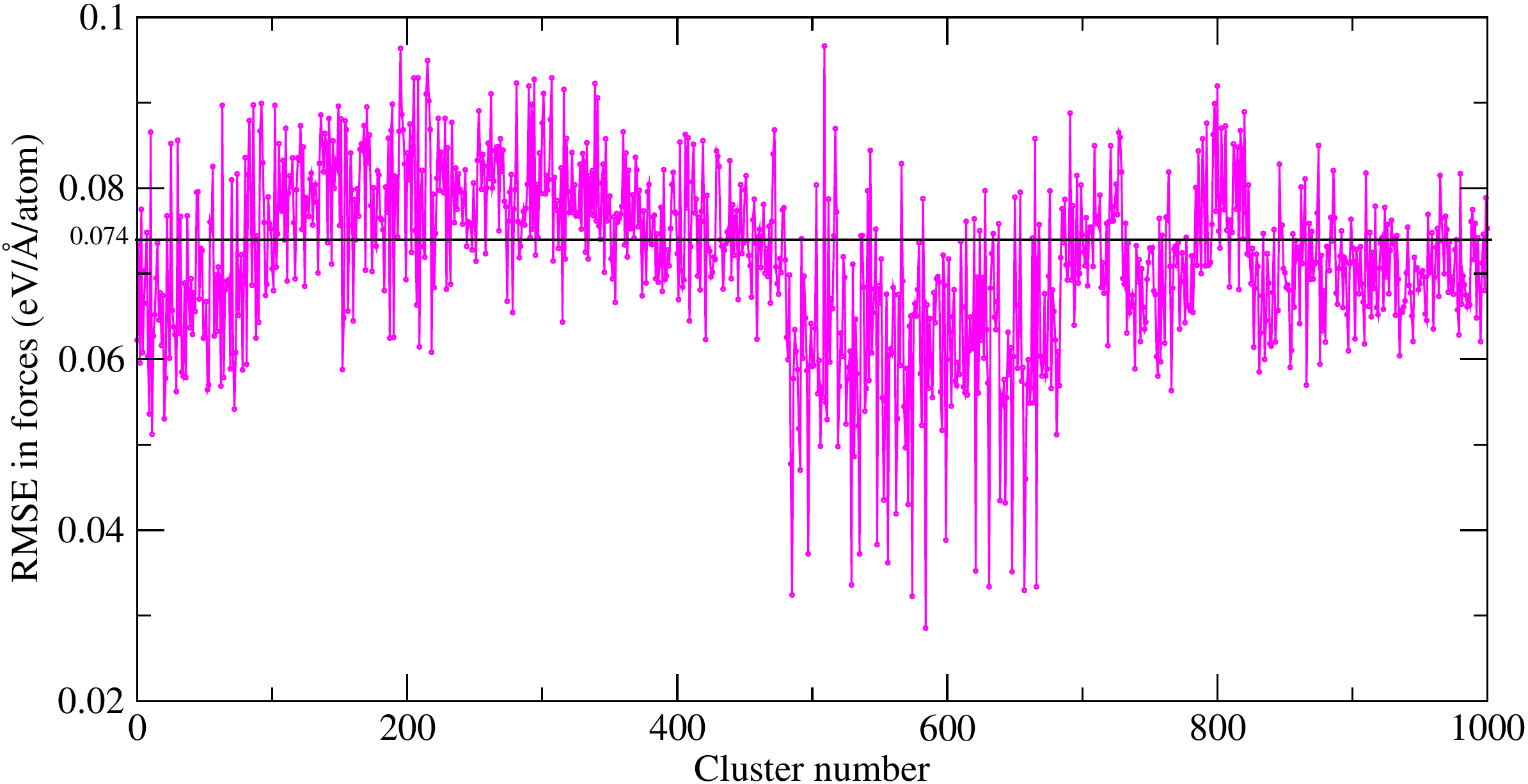}
\caption{A plot of RMSE of forces for $(AgAu)_{55}$ - $(AgAu)_{147}$}
\label{comp_dftannF_agau}
\end{figure}

\begin{figure}[htp]
\centering
\includegraphics[width=2.2in,height=2.1in]{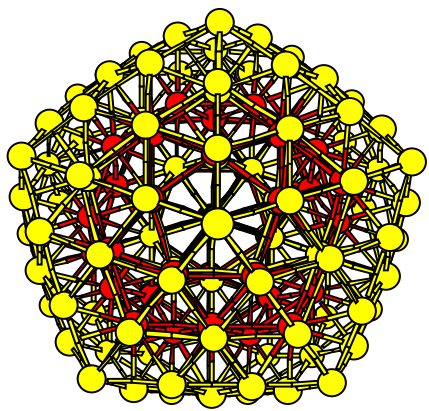}
\caption{Initial structure of $Ag_{35}Au_{112}$ for MDS}
\label{initalloystr}
\end{figure}

\begin{figure}[htp]
\centering
\includegraphics[width=2.2in,height=2.1in]{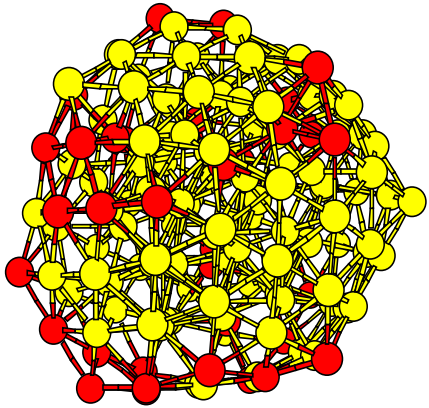}
\caption{Structure of $Ag_{35}Au_{112}$ obtained at 600 K}
\label{600Kstr}
\end{figure}

On fitting the energy and forces with our proposed approach, we got an average root mean square error (RMSE) of 5.9 meV/atom for energy of a cluster and 74 meV/\AA /atom for atomic forces. In order to verify the prediction potential of the weights, we compared the DFT and ANN energies for a small set of 500 clusters and have plotted in Fig. \ref{comp_dftann_agau}. We have also plotted the RMSE of forces for a set of 1000 clusters as shown in Fig. \ref{comp_dftannF_agau}. It is observed from the plot that a mojority of clusters lie below the average RMSE of forces. To validate the efficiency of the fitted energy and forces for the bimetallic system, we primarily focussed on geometry optimizations and MDS. We took an initial structure of $Ag_{35}Au_{112}$ consisting of three layers of atoms arranged in icosahedron geometry with silver atoms occupying the middle core and rest of the structure containing the gold atoms as shown in Fig. \ref{initalloystr}. We chose such an arrangement as its already studied\cite{agau_jcp,agau147_2} that in Au rich nanoalloys, gold atoms occupy surface and core atoms. We ran MDS at different temperatures - 300 K, 400 K, 500 K and 600 K. A time step of 1 \textit{fs} was used and the simulations were ran for a total time of 1 \textit{ns}. Since $(AgAu)_{147}$ is a large system, \textit{ab initio} MDS have not been performed yet. Various studies\cite{agau147_1,agau147_2} have been done using empirical potentials but they lack the QM accuracy. The optical absorption spectra\cite{agau147_3_dft} has been studied using first priciples but dynamics has not been explored. Using ANN parameters fitted to QM data, we have unravelled the dynamics of $Ag_{35}Au_{112}$ and have got quiet interesting results. 
It has been observed that with time the icosahedron geometry is completely dissolved and there is a huge variation in the atomic arrangement. At 300 K, the initial structure is maintained for a time of 13 \textit{ps}, and then silver atoms start to move towards surface. The gold atoms are too in a state of continuous rotational and vibrational motion but the geometry of inner core is maintained as such. As the simulation time is progressed, the geometry of the inner core changes from 13 gold atoms to 10 gold atoms and the middle core atomic arrangement alters from 42 atoms to 37 atoms comprising of both silver and gold atoms. Overall the surface atoms increases from 92 to around 100 atoms. This atomic arrangement of 100-37-10 is almost similar to the atomic arrangement in the global minimum structure of $Au_{147}$ as recently studied by Jindal et al.\cite{sphharmjcp,mybispec} It shows that icosahedron geometry is not favoured for pure gold or gold rich clusters. As we moved to simulations at higher temperatures, we observed that more of silver atoms are moving from middle core to the surface. One of the interesting observation was that silver atoms never entered the inner core. They either occupied the middle core or lied on the surface. Also, at a temperature of 600 K, almost all the silver atoms enriched the surface which is in accordance with the results for 24 $\%$ composition of Ag in $(AgAu)_{55}$ as published by Chiriki et al.\citep{agau_jcp} The structure obtained at 600 K is shown in Fig. \ref{600Kstr} in which the surface enriched with Ag can be seen.

\begin{figure}[!htb]
\centering
\subfigure[]
{
  \includegraphics[width=2.8in,height=1.9in]{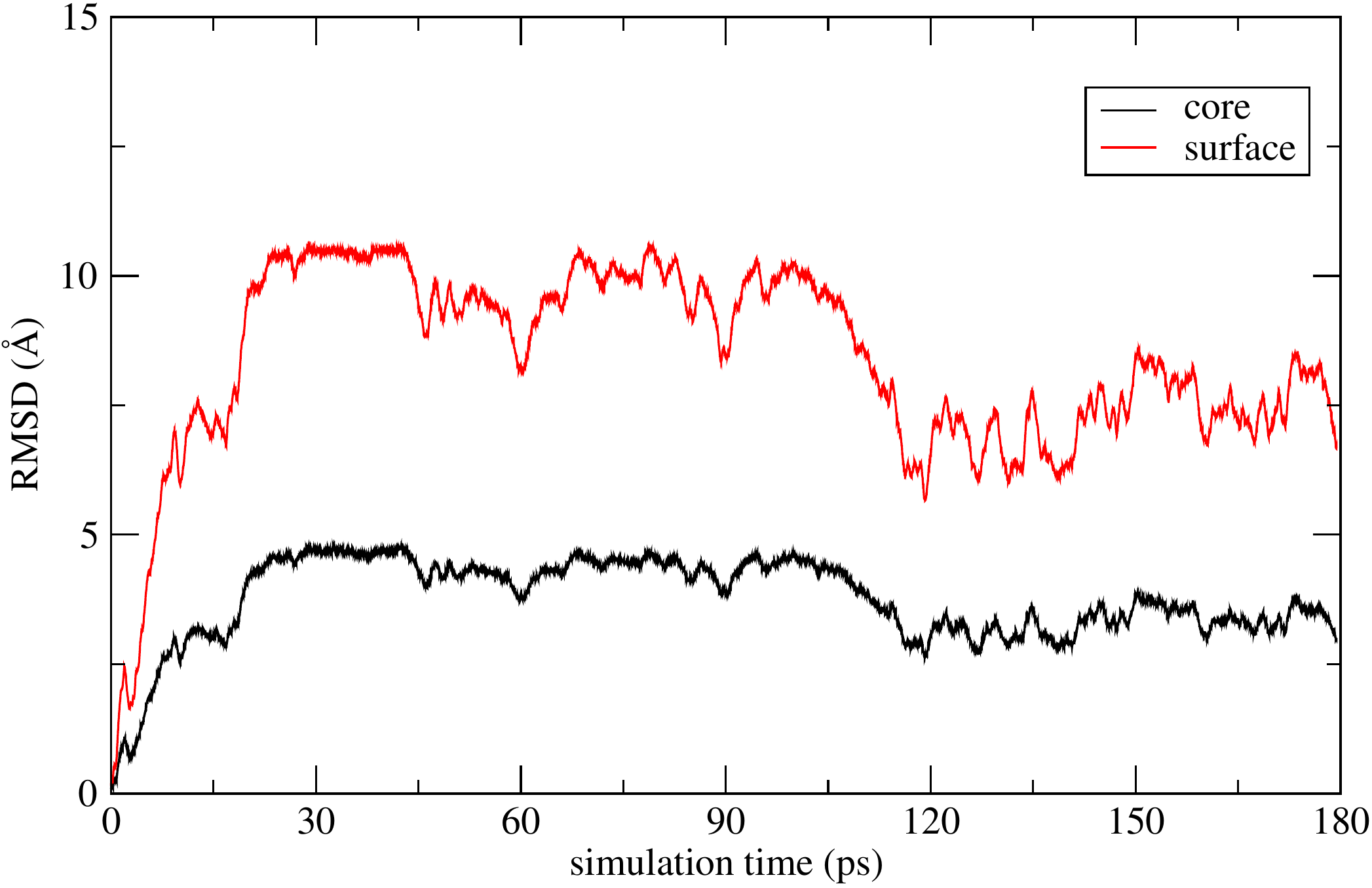}\label{rmsd300}
  }
\subfigure[]
{
  \includegraphics[width=2.8in,height=1.9in]{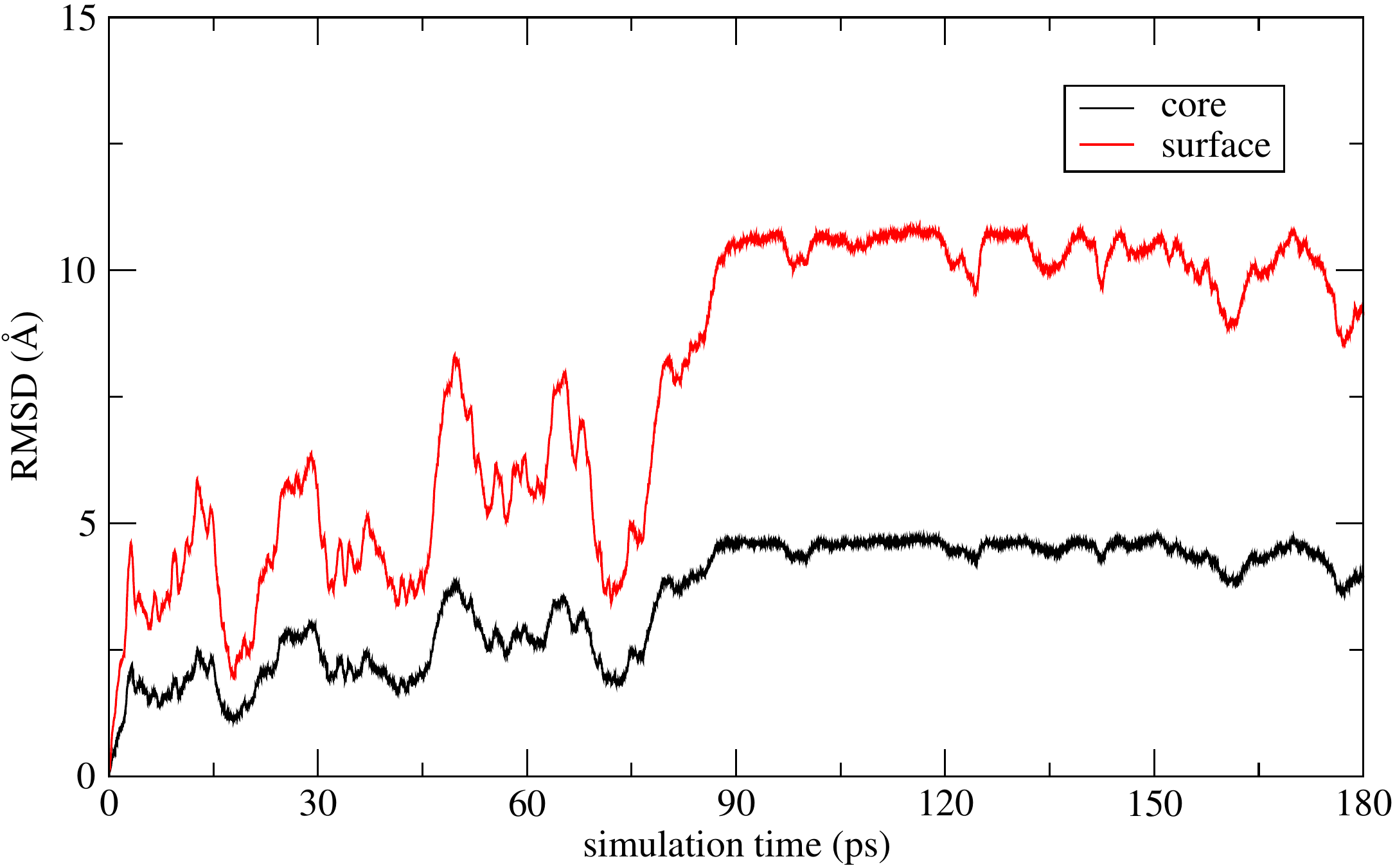}\label{rmsd400}
  } \\
\subfigure[]
{
  \includegraphics[width=2.8in,height=1.9in]{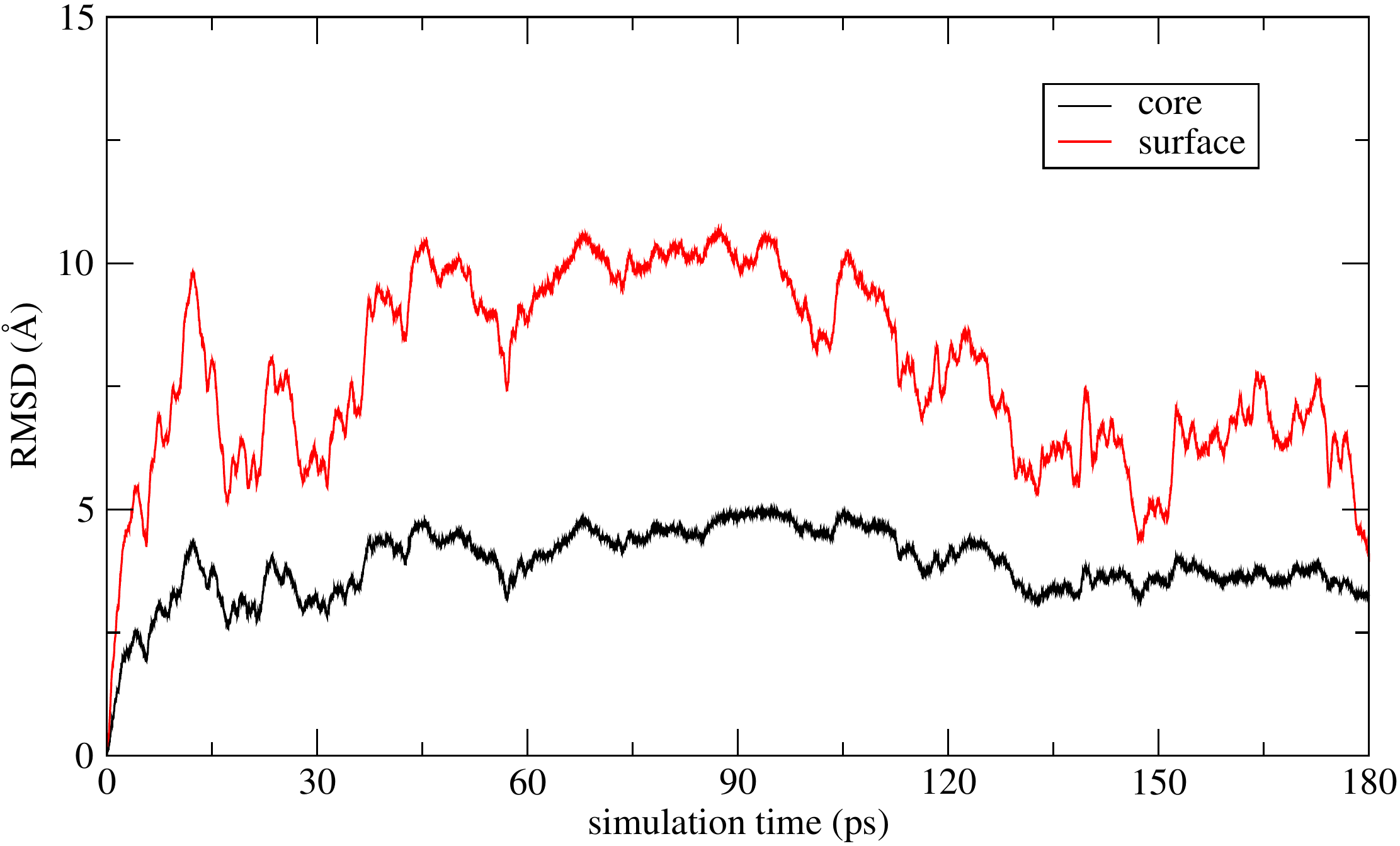}\label{rmsd500}
  }
  \subfigure[]
{
  \includegraphics[width=2.8in,height=1.9in]{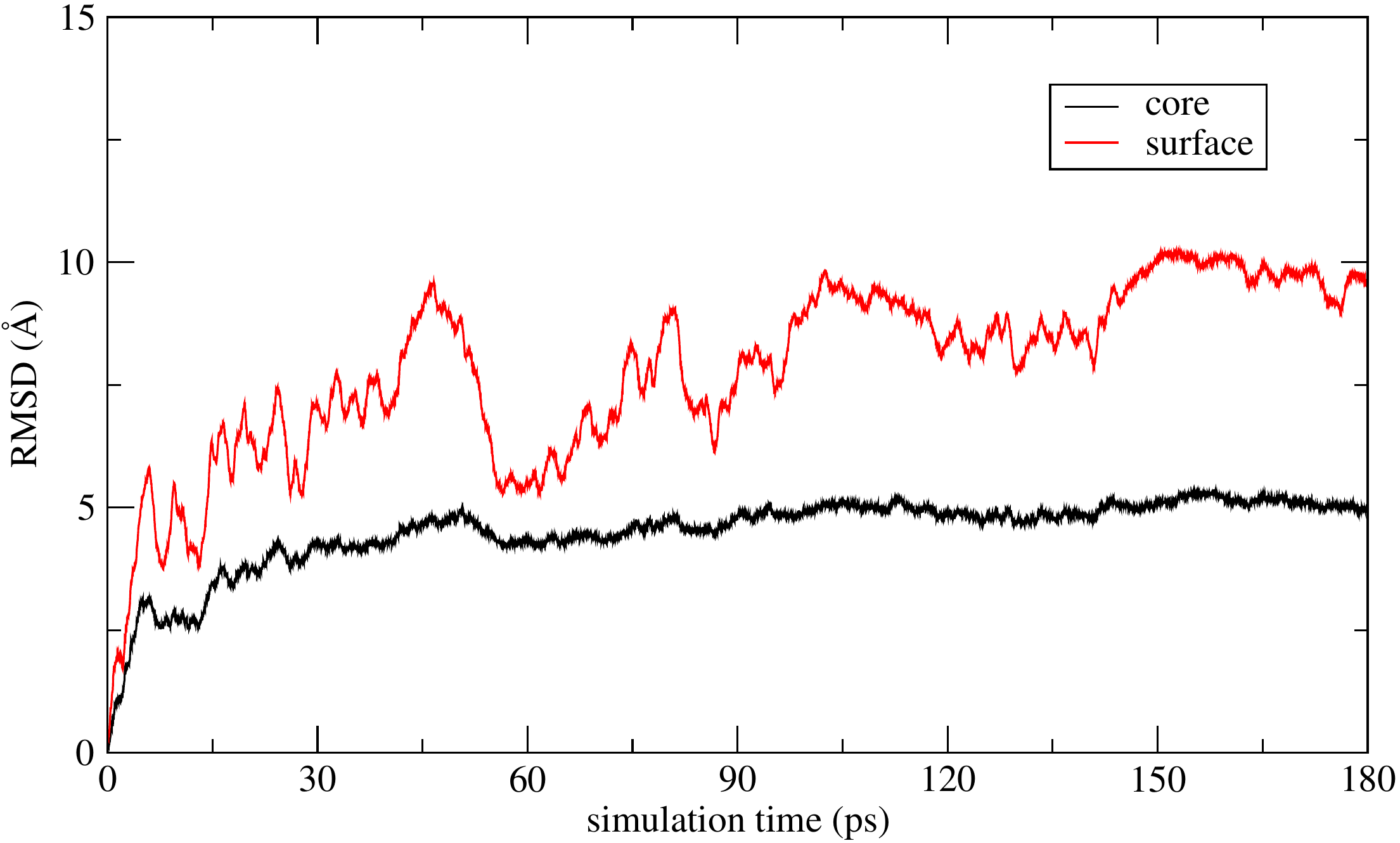}\label{rmsd600}
  }
\caption{RMSD plots for MDS of $Ag_{35}Au_{112}$ at (a) 300 K, (b) 400 K, (c) 500 K, (d) 600 K}\label{rmsdplots_agau}
\end{figure}

We have plotted root mean square distance\cite{gold_flux} (RMSD) as the order parameter to infer the movements in the cluster with simulation time. It calculates the average distance an atom has moved from the initial structure. For a structure at a given time, a sum over all the atomic movements is taken and divided by the total number of atoms in the structure.
\begin{equation}\label{rmsdeqn}
\text{RMSD} = \sqrt{\frac{\sum_{v=1}^{\text{atoms}}(x_v-x_v^s)^2+(y_v-y_v^s)^2+(z_v-z_v^s)^2}{N}} 
\end{equation}
In Eq. \ref{rmsdeqn}, $x_v$, $y_v$ and $z_v$ are the Cartesian coordinates of the initial structure and $x_v^s$, $y_v^s$ and $z_v^s$ are the Cartesian coordinates of the structure at a given simulation time. The RMSD plots at 300 K, 400 K, 500 K and 600 K is shown in Fig. \ref{rmsdplots_agau}. It can be inferred that both the core atoms and the surface atoms of $Ag_{35}Au_{112}$ undergo a lot of movements thus making it a highly fluxional cluster. With the increase in simulation time, it is observed in all the temperatures that both surface and the core atoms try to attain the geometry as of initial structure, but since that structure is not stable, the geometry changes to a more stable arrangement of atoms. 

\begin{figure}[htp]
\centering
\includegraphics[width=4.2in,height=2.6in]{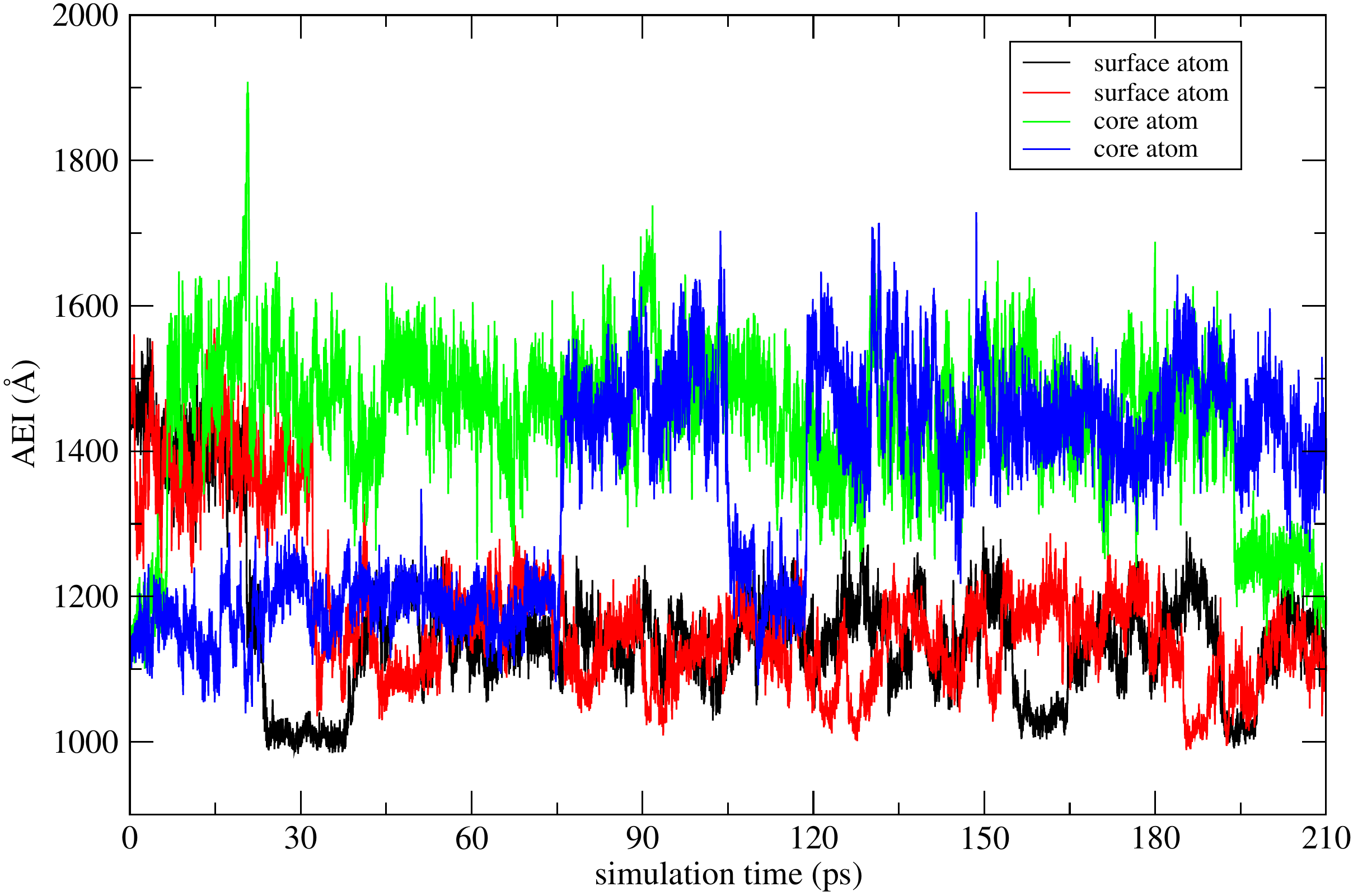}
\caption{The AEI of two core atoms and two surface atoms in $Ag_{35}Au_{112}$ throughout the MDS at 600 K}
\label{AEI_600}
\end{figure}

In order to show the inter mixing of atoms between surface and core at 600 K, we have plotted atomic equivalence indices (AEI).\cite{AEI} 
\begin{equation}
\text{AEI}^{h} = \sum_{k}|\vv{R_h}(t)-\vv{R_k}(t)|
\end{equation}
Here, $\vv{R_h}(t)$ is the position vector of an atom $h$ at a particular simulation time $t$. Its a very sensitive indicator and maps even the tiny movements throughout the simulations. Since its a $147$ atom cluster, therefore it is not possible to plot the AEI for all the atoms. Therefore, we selected 4 atoms from the structure in which two are the core atoms and other two are surface atoms. The plot is shown in Fig. \ref{AEI_600}. There is a continuous movement of the core atom to the surface and back to the core, as seen by the blue colored curve in Fig. \ref{AEI_600}. The surface atoms are moving but not entering the core as observed from the red and the black colored curves in Fig. \ref{AEI_600}.

\begin{figure}[!htb]
\centering
\subfigure[]
{
  \includegraphics[width=1.9in,height=1.30in]{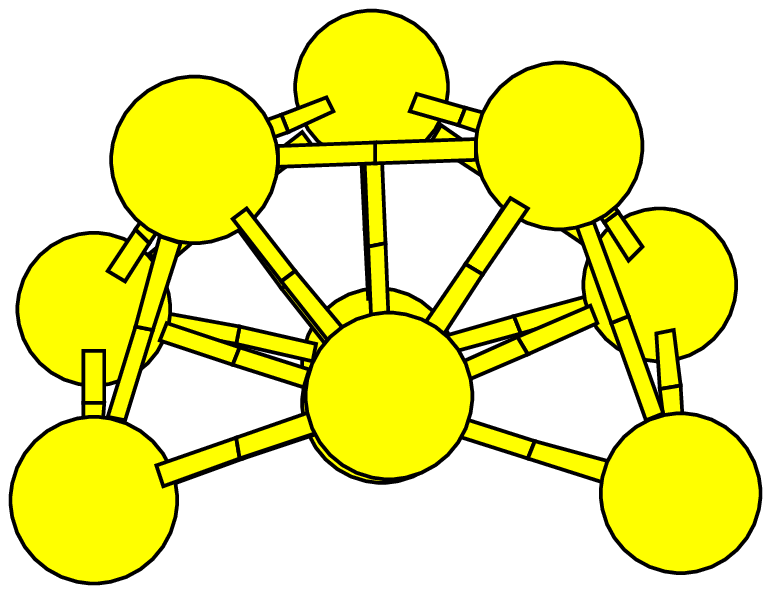}\label{10atomside}
  }
\subfigure[]
{
  \includegraphics[width=1.9in,height=1.30in]{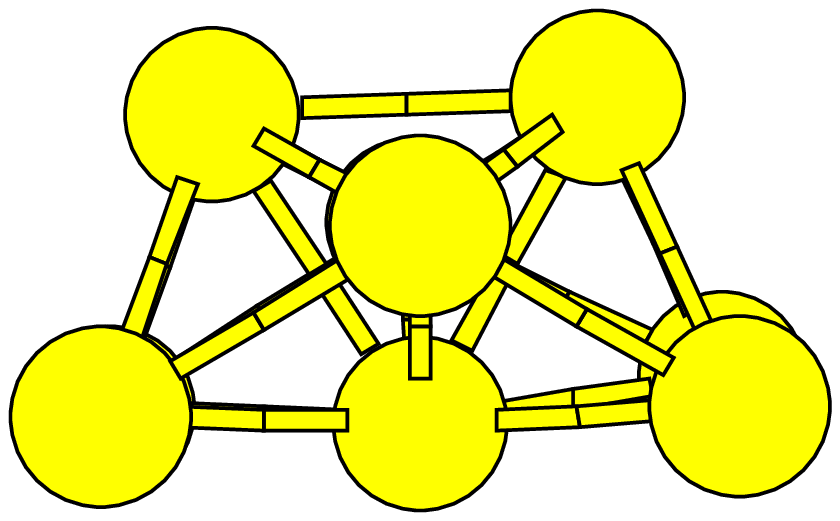}\label{9atomside}
  } \\
\subfigure[]
{
  \includegraphics[width=1.9in,height=1.5in]{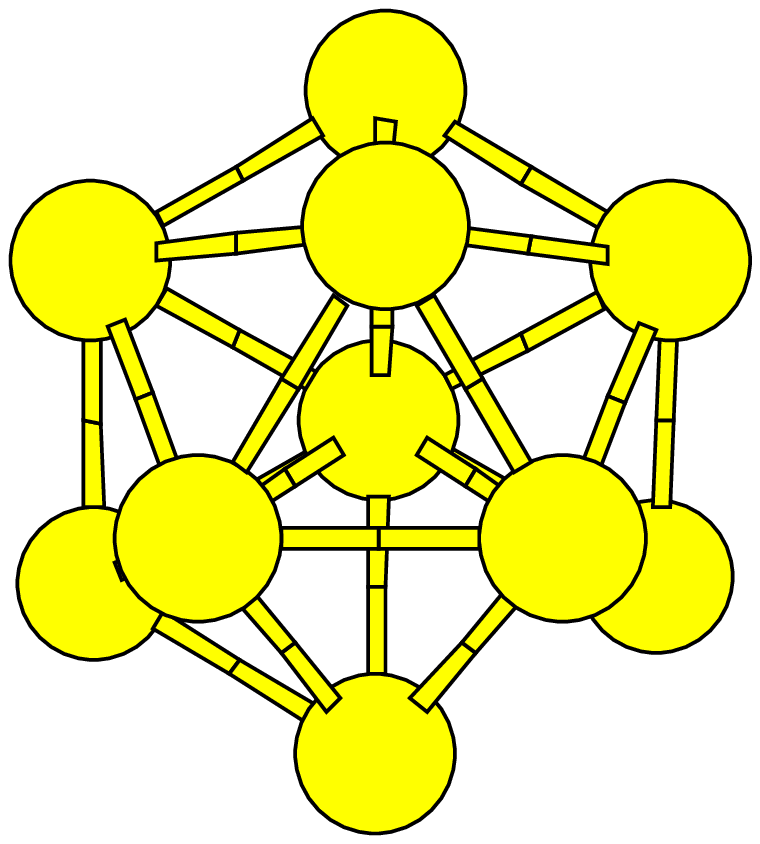}\label{10atomtop}
  }
  \subfigure[]
{
  \includegraphics[width=1.9in,height=1.5in]{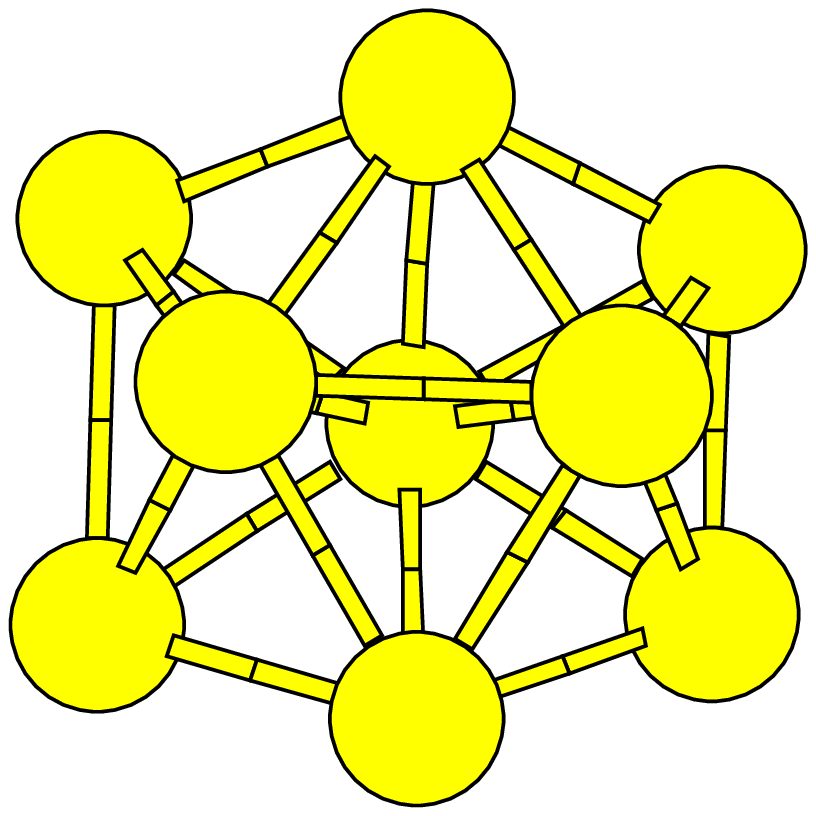}\label{9atomtop}
  }
\caption{The inner core geometry for $Ag_{35}Au_{112}$ (a) 10 atom inner core (side view), (b) 9 atom inner core (side view), (c) 10 atom inner core (top view), (d) 9 atom inner core (top view)}\label{innercoregeom}
\end{figure}

\begin{figure}[!htb]
\centering
\subfigure[]
{
  \includegraphics[width=2.2in,height=2.1in]{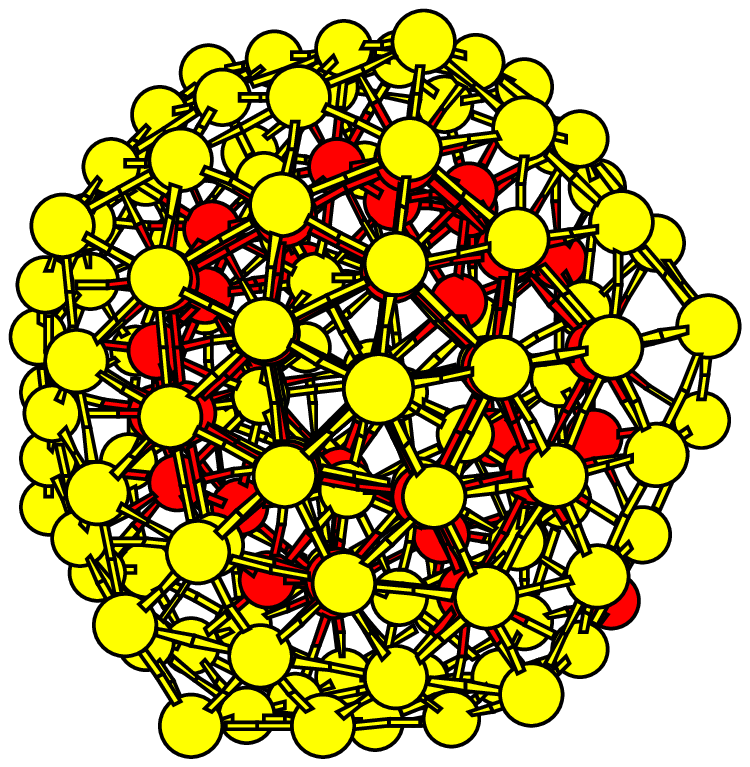}\label{lowestisom}
  }
\subfigure[]
{
  \includegraphics[width=2.2in,height=2.1in]{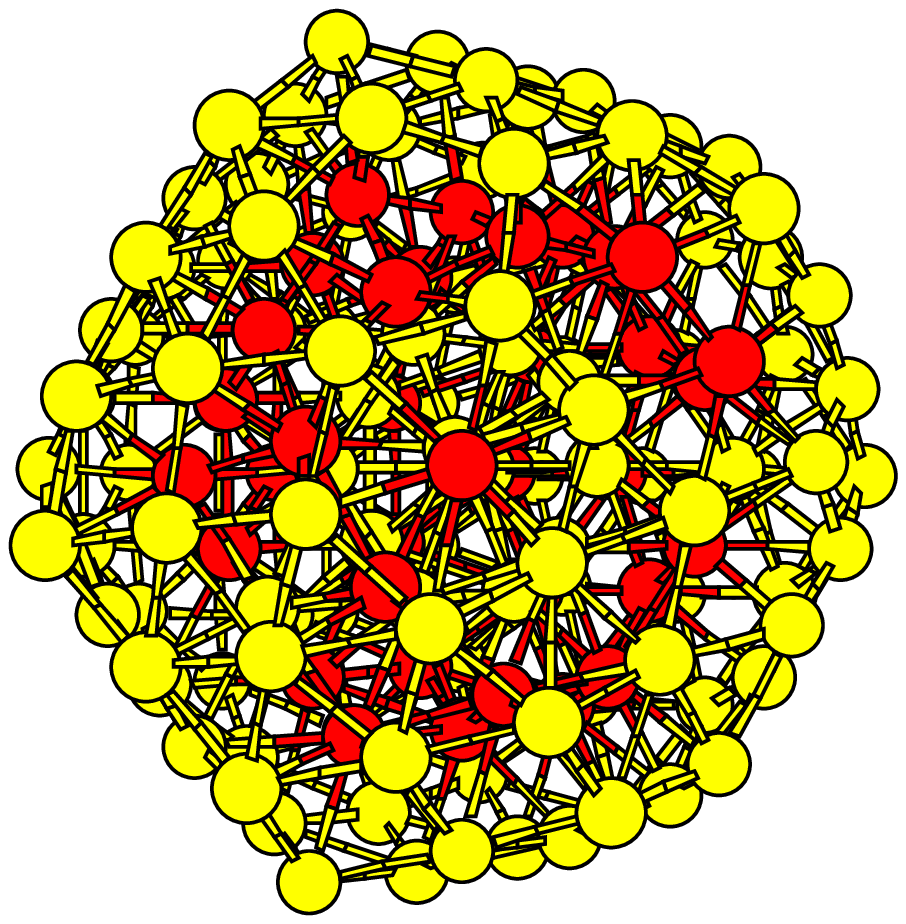}\label{iso40}
  } 
\caption{(a)The lowest energy structure quenched for $Ag_{35}Au_{112}$, (b) The structure consisting of 9 atom inner core and lying 0.39 eV higher in energy than lowest energy structure of $Ag_{35}Au_{112}$.}\label{lowestisomers}
\end{figure}

After running the MDS for 1 \textit{ns}, we collected the local minima structures from trajectories obtained at different temperatures. We ran geometry optimizations using Broyden - Fletcher - Goldfarb - Shanno (BFGS) algorithm\cite{lbfgs} and got some symmetric inner core geometries as shown in Fig. \ref{innercoregeom}. The lowest energy isomer that we quenched from the MDS trajectories is shown in Fig. \ref{lowestisom}. It contained 10 atoms in the inner core, 37 atoms in the middle core and 100 atoms on the surface. From the initial structure of MDS, 5 silver atoms moved to the surface forming the lowest energy isomer. Another isomer with 9 atoms in the inner core, 36 atoms in the middle core and 102 atoms on the surface is shown in Fig. \ref{iso40}. There was a difference of 0.39 eV between the two isomers, showing a possibility of large number of fluxional isomers for $Ag_{35}Au_{112}$. Overall, a cage like structure makes the foundation of gold rich $Ag_{35}Au_{112}$ alloy. Our method can be further extended to study global optimizations in such alloys. Also, icosahedron geometry is not the stable isomer for 147 atom configuration.
\subsection{Thiol protected gold nanoclusters: Study of $Au_{68}(SH)_{32}$}
\begin{figure}[htp]
\centering
\includegraphics[width=4.2in,height=2.6in]{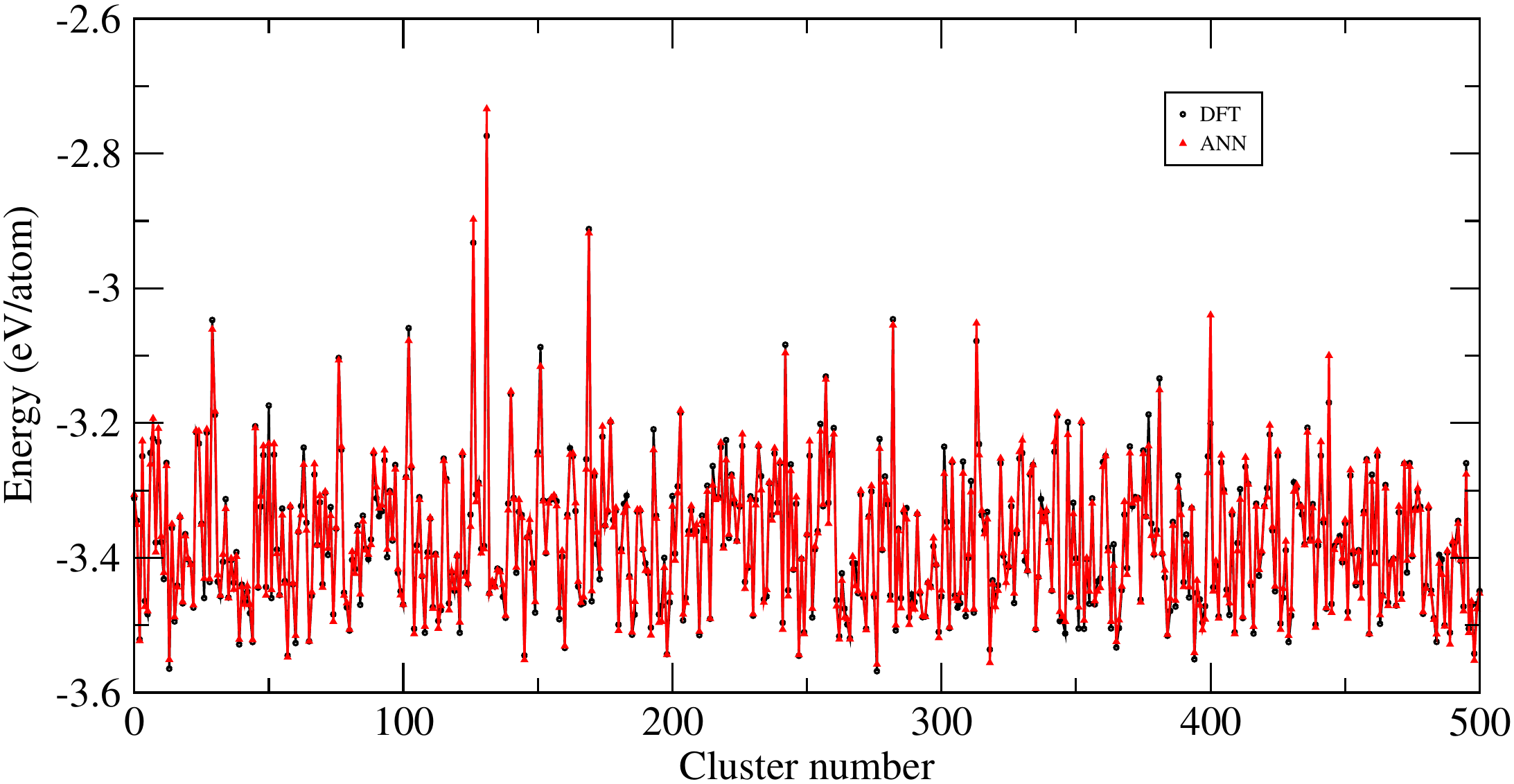}
\caption{Comparison of DFT and ANN predicted energies for $Au_m(SH)_n$ clusters where $m$ varies from 13 to 38 and $n$ varies from 6 to 24}
\label{comp_dftann_aush}
\end{figure}

\begin{figure}[htp]
\centering
\includegraphics[width=4.2in,height=2.6in]{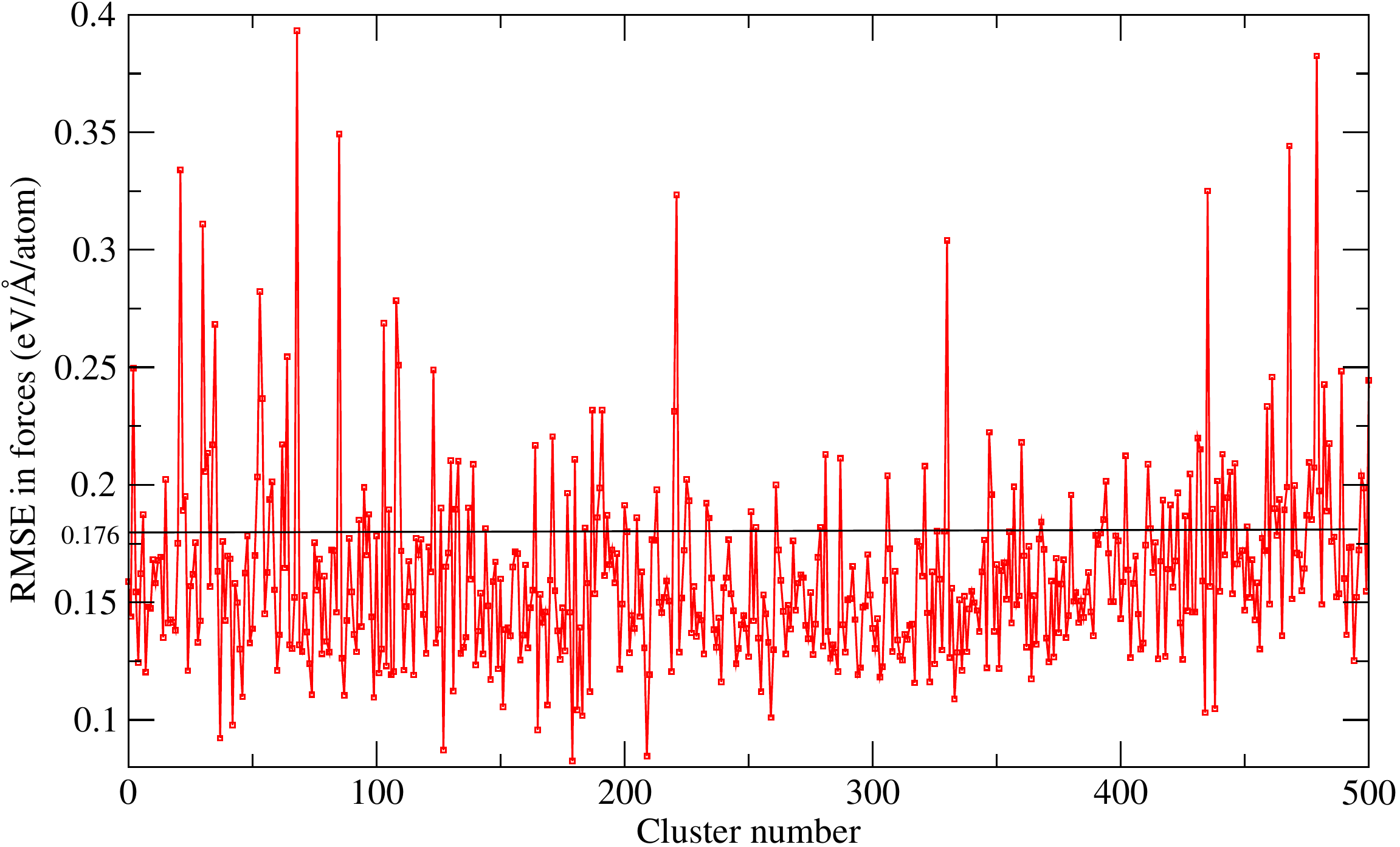}
\caption{Plot of RMSE of forces for $Au_m(SH)_n$ clusters where $m$ varies from 13 to 38 and $n$ varies from 6 to 24}
\label{comp_dftFann_aush}
\end{figure}

\begin{figure}[!htb]
\centering
\subfigure[]
{
  \includegraphics[width=2.2in,height=2.1in]{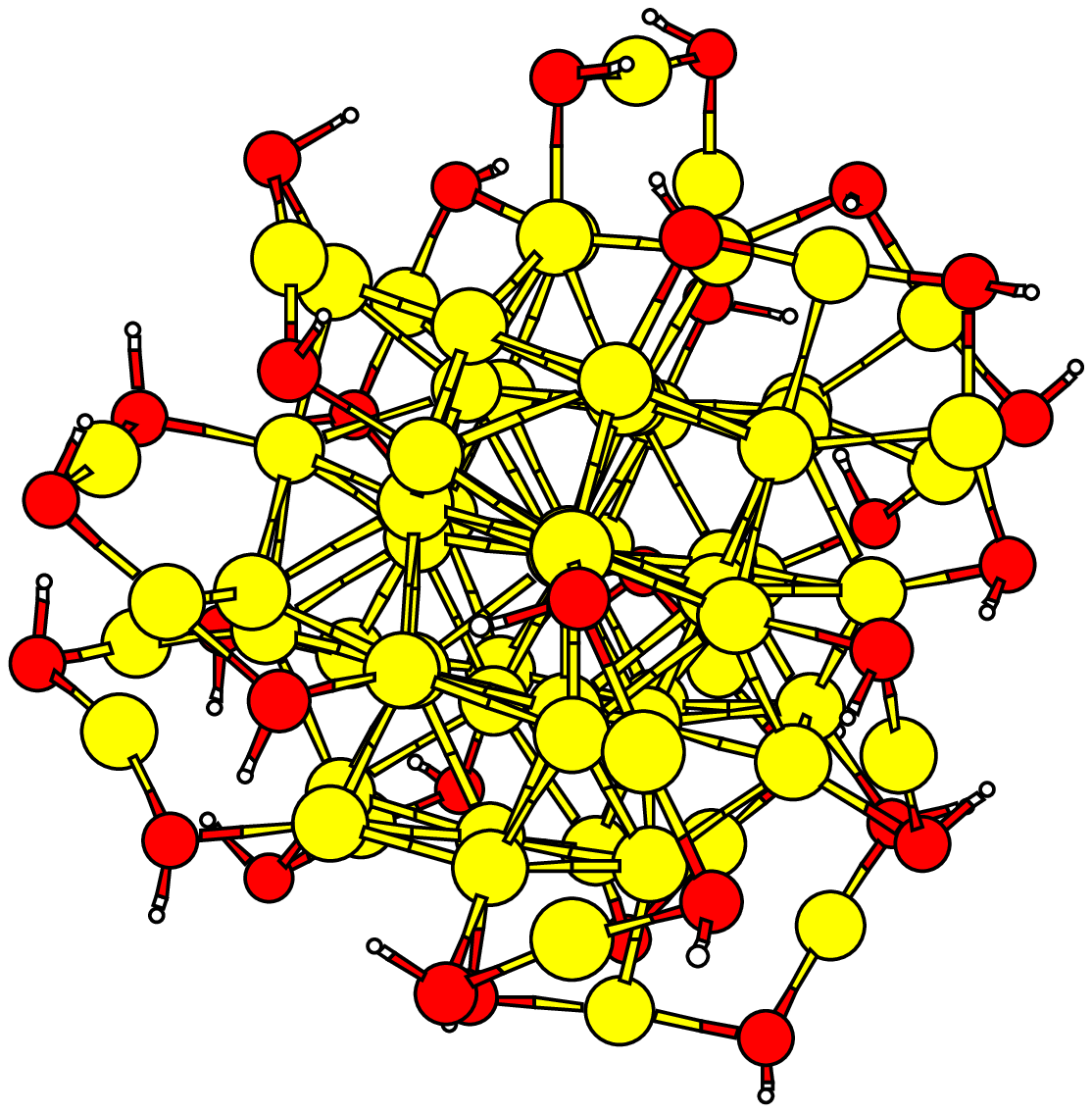}\label{gm_au68}
  }
\subfigure[]
{
  \includegraphics[width=2.2in,height=2.1in]{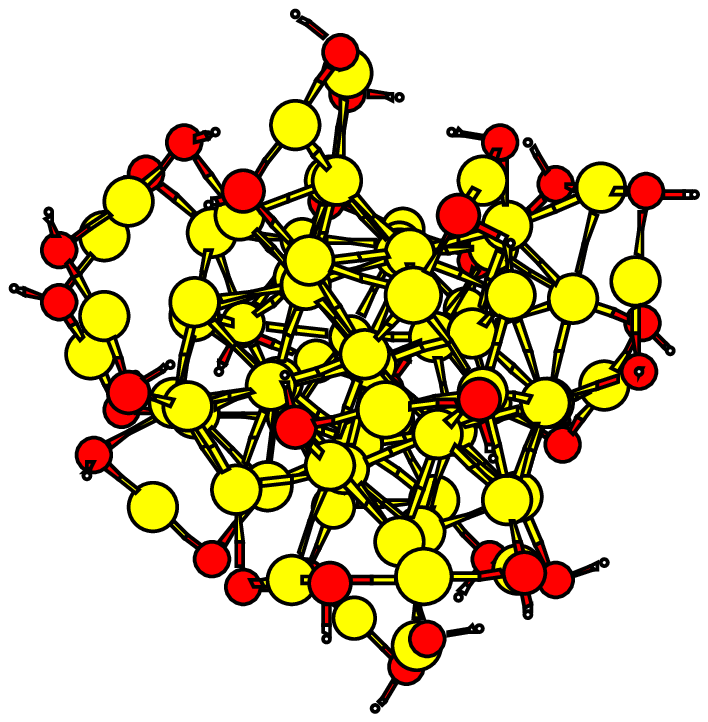}\label{locmin1}
  } 
  \subfigure[]
{
  \includegraphics[width=2.2in,height=2.1in]{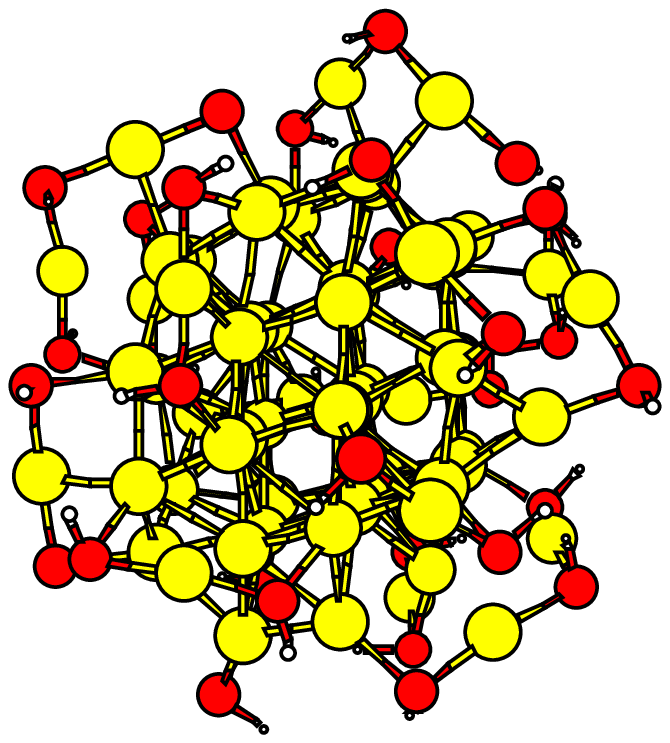}\label{locmin2}
  }
\subfigure[]
{
  \includegraphics[width=2.2in,height=2.1in]{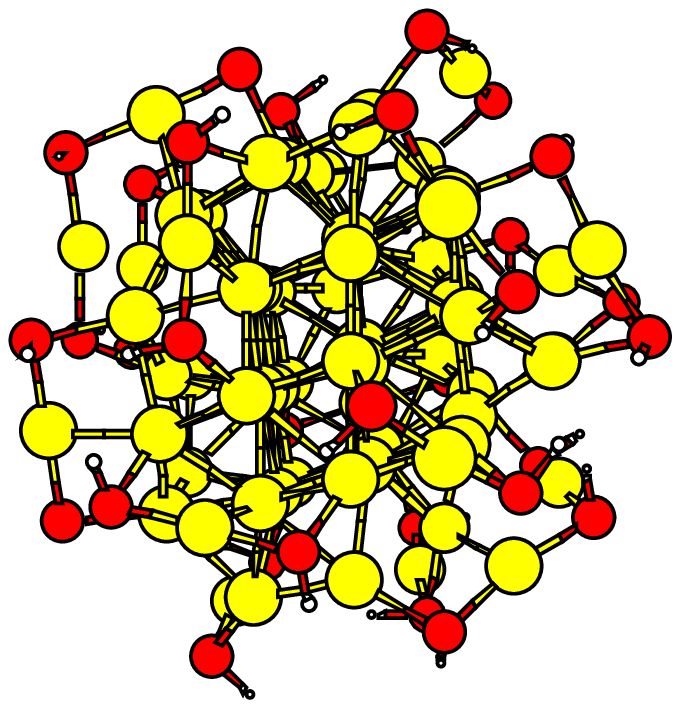}\label{locmin3}
  } 
\caption{Optimized structures of $Au_{68}(SH)_{32}$ similar to that obtained by Xu et al.\cite{Au68_zeng} (a)The global minimum, (b),(c) and (d) are the local minimas.}\label{optisomau68}
\end{figure}

We selected thiol protected gold nanoclusters for validation of our proposed method due to the increased complexity in the structure. The gold, sulphur and hydrogen atoms have different valence electrons and orbital configuration which leads to different patterns of bonding between each other. Hence, using the descriptors, the accurate prediction of the forces is a challenge. So, on fitting the dataset consisting of clusters from $Au_{13}(SH)_{6}$-$Au_{38}(SH)_{24}$, we got an average RMSE of 8.6 meV/atom for energy of a cluster and 176 meV/\AA /atom for atomic forces. It should be noted that the reason for getting a higher RMSE as compared to gold-silver nanoalloys is the huge variations in thiol protected gold clusters which makes it difficult to fit. We validated the weights for a set of 500 clusters consisting of $Au_m(SH)_n$ clusters ($m$ : 13 - 38 and $n$ : 6 - 24) and compared their energies with the DFT predicted energies as shown in Fig. \ref{comp_dftann_aush}. We also plotted the RMSE of forces for these clusters as shown in Fig. \ref{comp_dftFann_aush}. Here, we can see that a lot of clusters have RMSE lower than the average RMSE (176 meV/\AA /atom), thus showing an accurate fitting of forces. We extrapolated our weights to study geometry optimization and dynamics of $Au_{68}(SH)_{32}$. On geometry optimization of the global minimum and the local minimas predicted for $Au_{68}(SH)_{32}$ by Xu et al.,\cite{Au68_zeng} we got the same structures as theirs which is shown in Fig. \ref{optisomau68}. This reflects that the atomic forces have fitted very well and have captured necessary bonding patterns between $Au$, $S$ and $H$. Also, the dataset $Au_{13}(SH)_{6}$-$Au_{38}(SH)_{24}$ forms a subset for the atomic environments of $Au_{68}(SH)_{32}$.

\begin{figure}[htp]
\centering
\includegraphics[width=3.1in,height=2.2in]{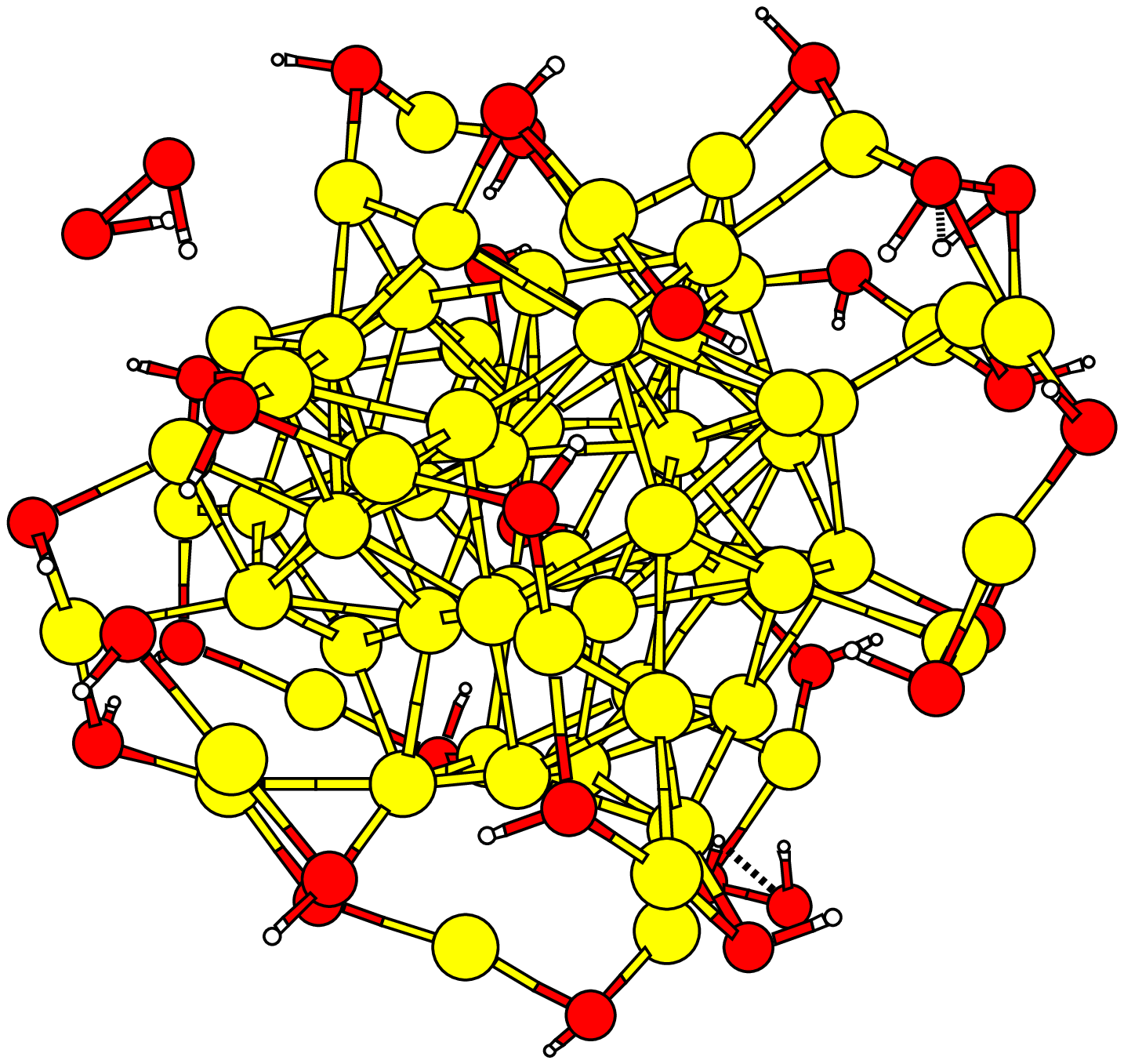}
\caption{Desorption of thiol group from Au in $Au_{68}(SH)_{32}$ at 300 K}
\label{desorbstr}
\end{figure}

\begin{figure}[htp]
\centering
\includegraphics[width=3.6in,height=2.4in]{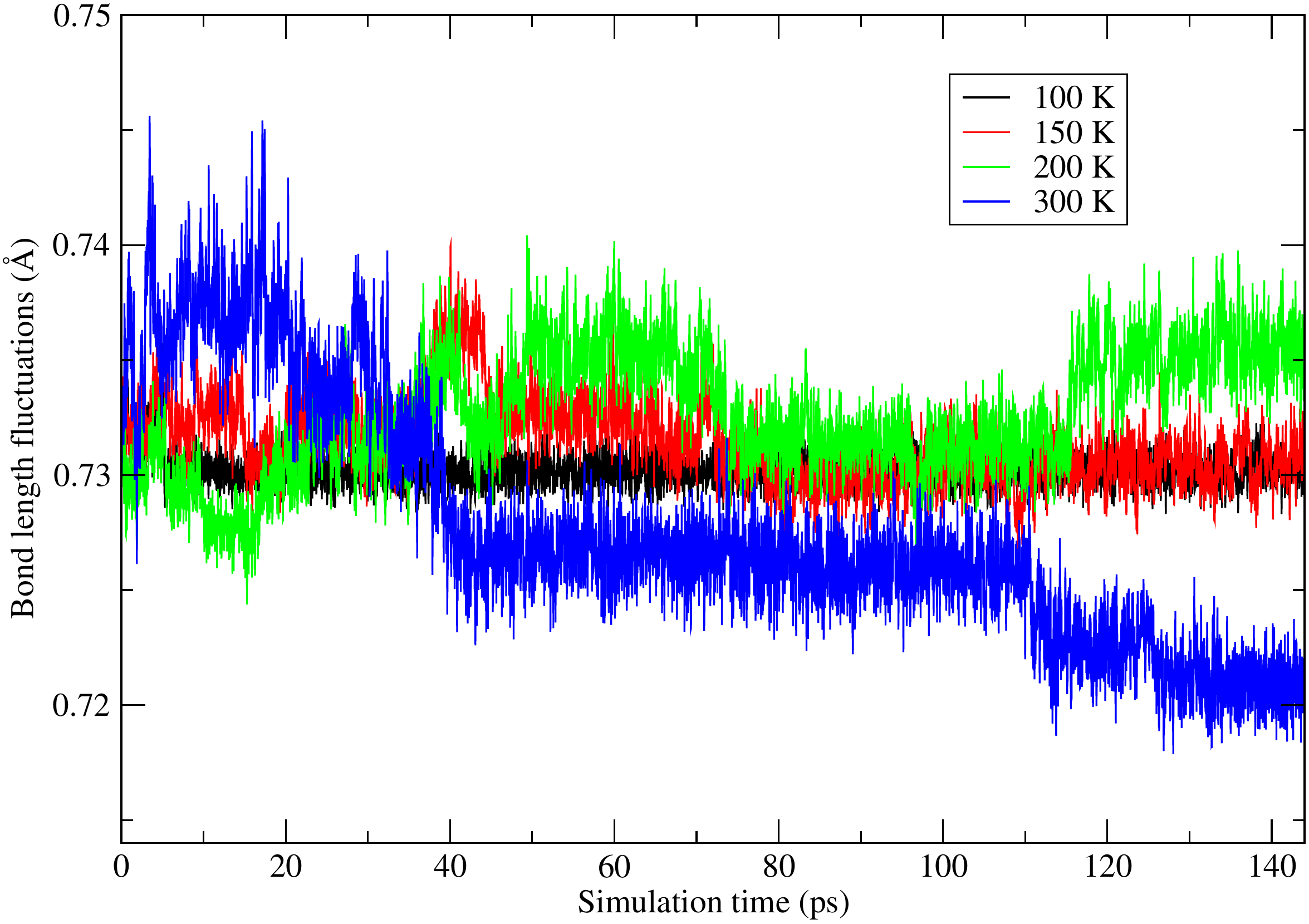}
\caption{Average bond length fluctuations between $S$ and staple-$Au$ during MDS at 100 K, 150 K, 200 K and 300 K of $Au_{68}(SH)_{32}$}
\label{avgblf_Au68SH32}
\end{figure}

Since, geometry optimization was achieved accurately, we applied the weights used for fitting forces to further run MDS. The MDS were run at temperatures - 100 K, 150 K, 200 K and 300 K at a time step of 0.1 \textit{fs} for a total time of 1 \textit{ns}. Running the dynamics at 300 K gave an important insight into structural stability of thiol protected gold nanoclusters. It was observed that thiol group undergoes desorption from the gold surface as shown in Fig. \ref{desorbstr}. This observation is in accordance with the work done by B\"{u}ttner et al.\cite{thiol_desorp} by using X-ray photoelectron spectroscopy for thiol passivated gold particles. To validate this observation, we plotted average bond length fluctuations\cite{bondlenflu} for the $S$ and staple - $Au$ bonds as shown in Fig. \ref{avgblf_Au68SH32}. The plot highlights that the S and staple- $Au$ are intact at a very low temperature of 100 K. But as the temperature is increased, the fluctuations increase in the beginning of the simulation and thus leads to a breakage in the bond between S and Au. This is clearly evident from the blue colored plot at 300 K shown in Fig. \ref{avgblf_Au68SH32}. Therefore, in order to maintain the protection of gold nanoclusters, they should be stabilised below 150 K such that thiol group does not desorb from gold surface.

\begin{figure}[!htb]
\centering
\subfigure[]
{
  \includegraphics[width=1.8in,height=1.40in]{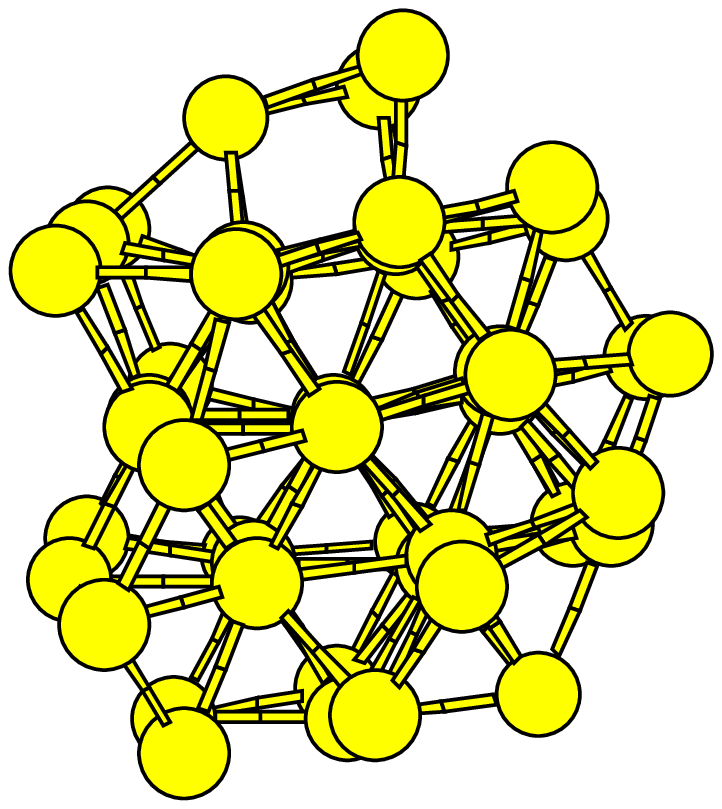}
  }
\subfigure[]
{
  \includegraphics[width=1.8in,height=1.40in]{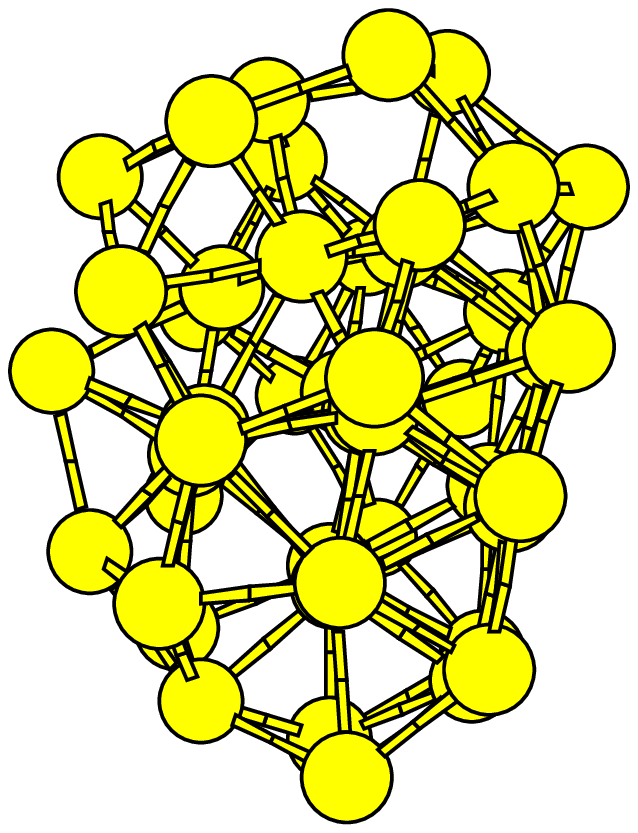}
  } \\
\subfigure[]
{
  \includegraphics[width=1.9in,height=1.5in]{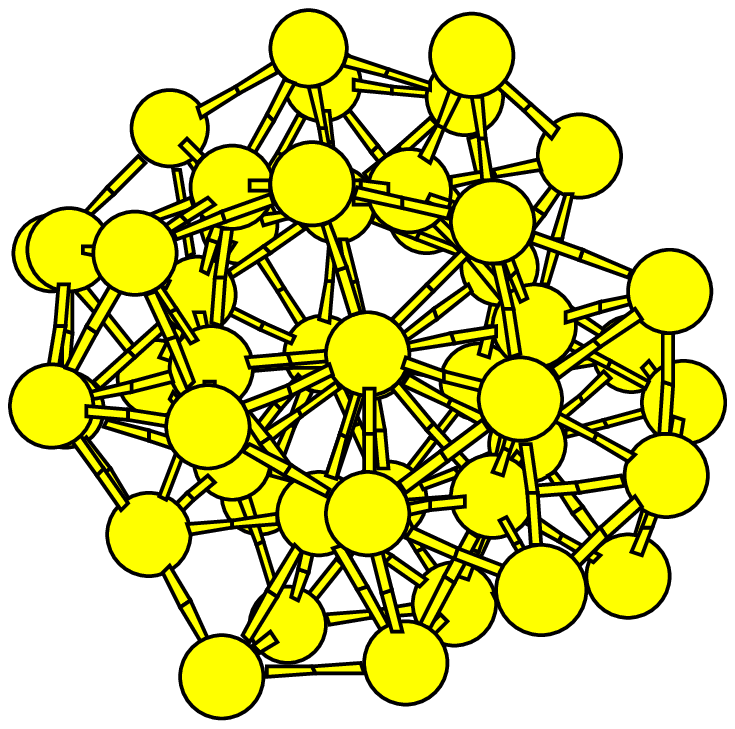}
  }
  \subfigure[]
{
  \includegraphics[width=1.9in,height=1.5in]{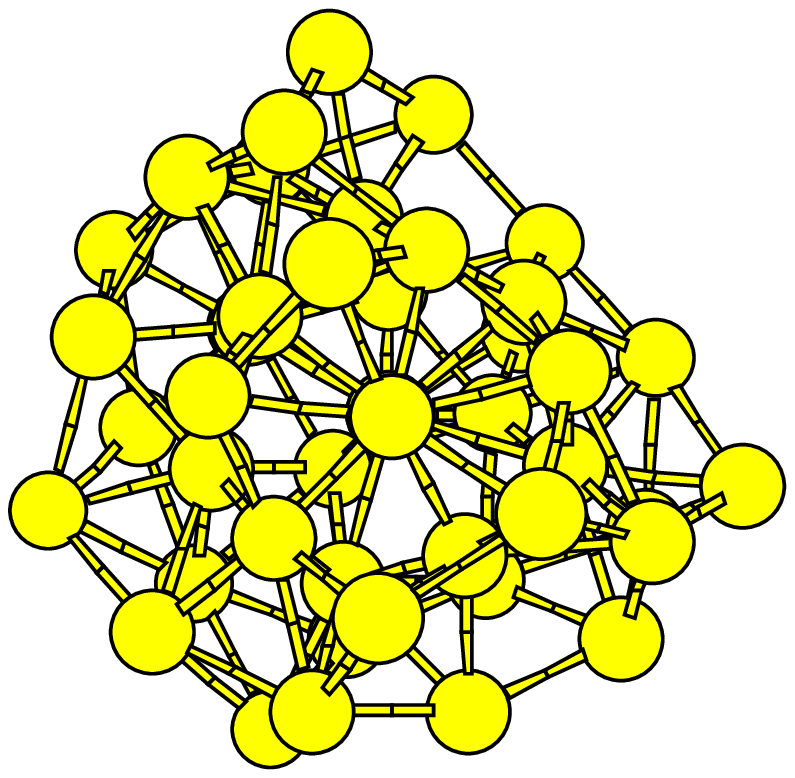}
  }
\caption{The core structures for $Au_{68}(SH)_{32}$}\label{coregeomau68}
\end{figure}

We also quenched the local minima structures from the MDS using BFGS algorithm. Some of the core geometries are shown in Fig. \ref{coregeomau68}.
\subsection{Extension of ANN to global optimizations (GO)}
After studying the ANN based MDS, we did global optimizations for gold-silver nanoalloys and thio- protected gold nanoclusters. We selected two clusters for gold-silver nanoalloys having 10{\%} ($Ag_5Au_{50}$) and 24{\%} ($Ag_{13}Au_{42}$) composition of $Ag$. For thio- protected gold nanoclusters, we selected $Au_{15}$ cluster protected with three different amounts of thiol units (8, 10 and 13). The GO were performed using basin hopping (BH)\cite{basihop} and MDS. The quenching of the structures was done using BFGS algorithm.\cite{lbfgs}

\begin{figure}[!htb]
\centering
\subfigure[]
{
  \includegraphics[width=1.8in,height=1.40in]{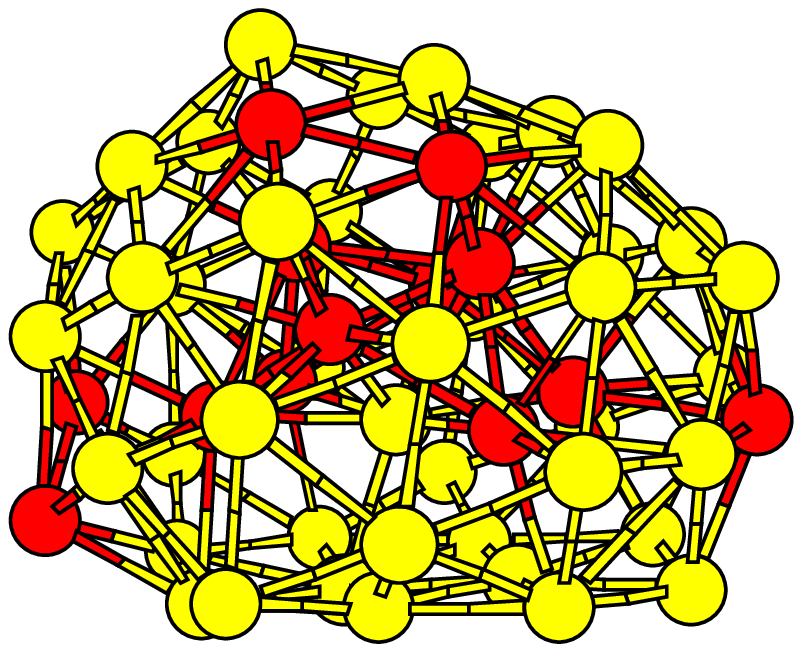}
  }
\subfigure[]
{
  \includegraphics[width=1.4in,height=1.0in]{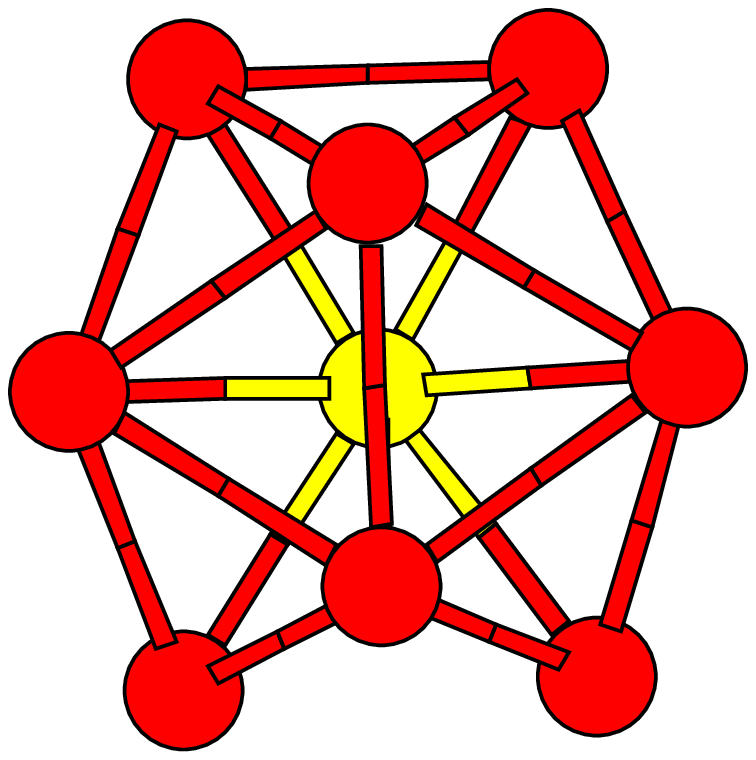}
  } 
\caption{(a) The global minimum structure of $Ag_{13}Au_{42}$, (b) The core structure of the global minimum of $Ag_{13}Au_{42}$}\label{gm_ag13au42}
\end{figure}

\begin{figure}[htp]
\centering
\includegraphics[width=3.6in,height=2.4in]{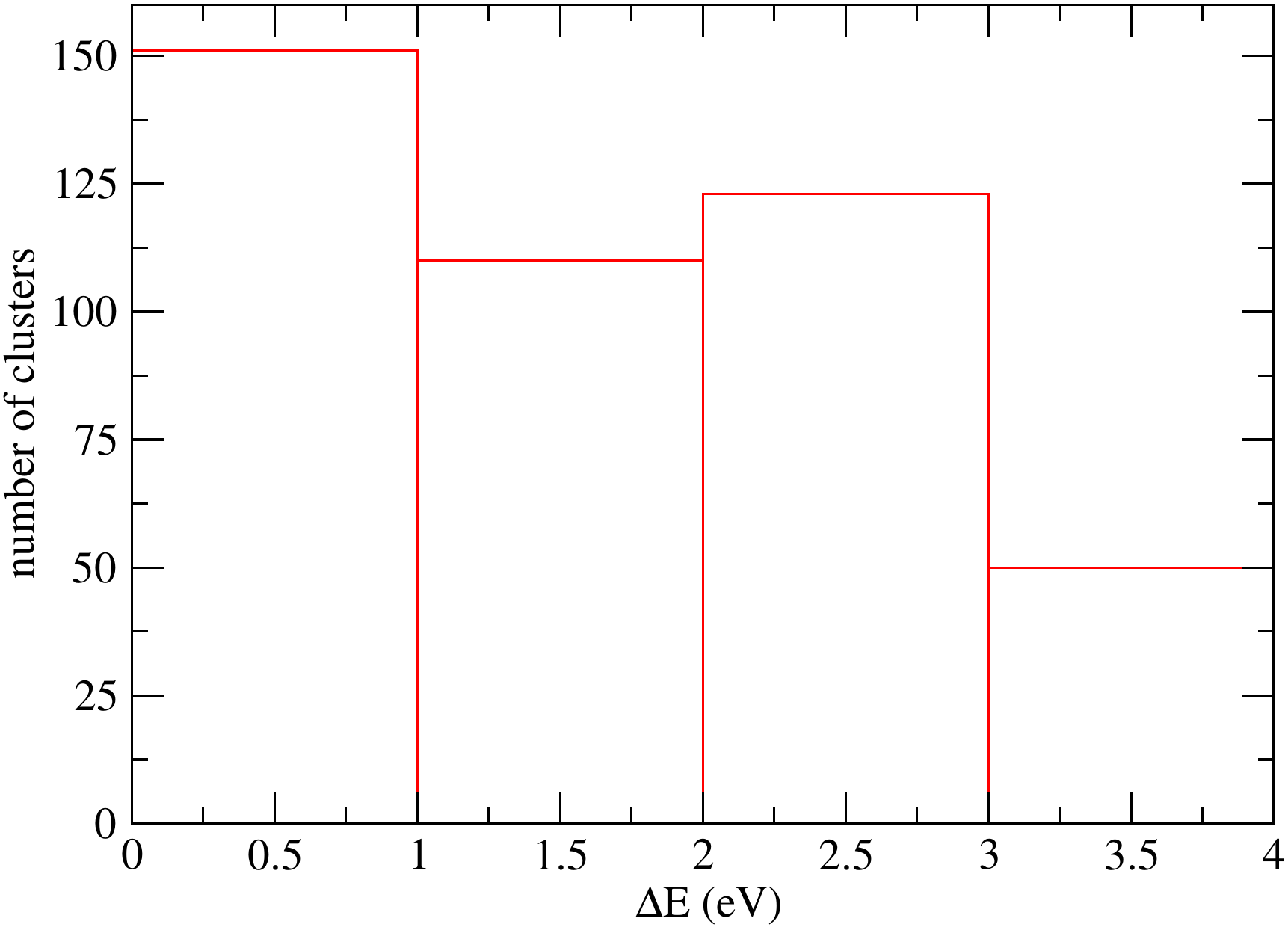}
\caption{A histogram showing the number of isomers found in an energy range from the GM of $Ag_{13}Au_{42}$}
\label{histog_ag13au42}
\end{figure}

\begin{figure}[!htb]
\centering
\subfigure[]
{
  \includegraphics[width=1.3in,height=1.2in]{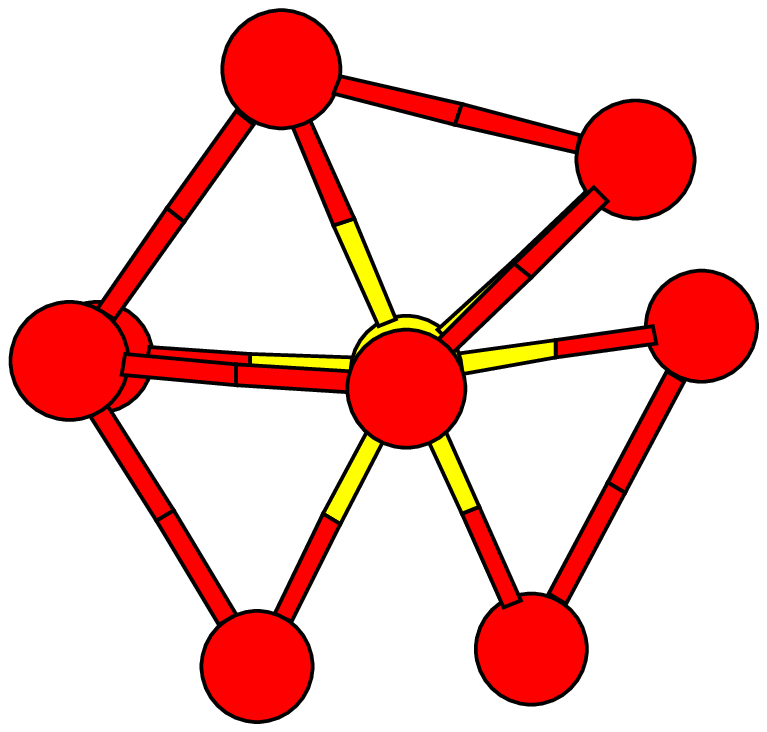}
  }
\subfigure[]
{
  \includegraphics[width=1.3in,height=1.1in]{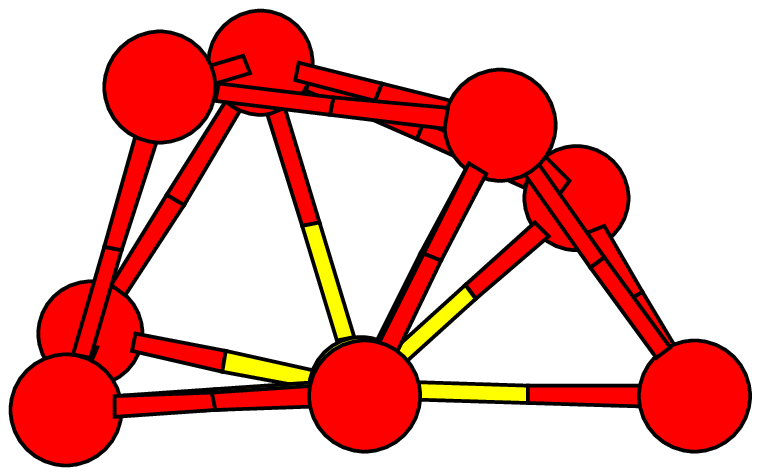}
  } \\
\subfigure[]
{
  \includegraphics[width=1.3in,height=1.2in]{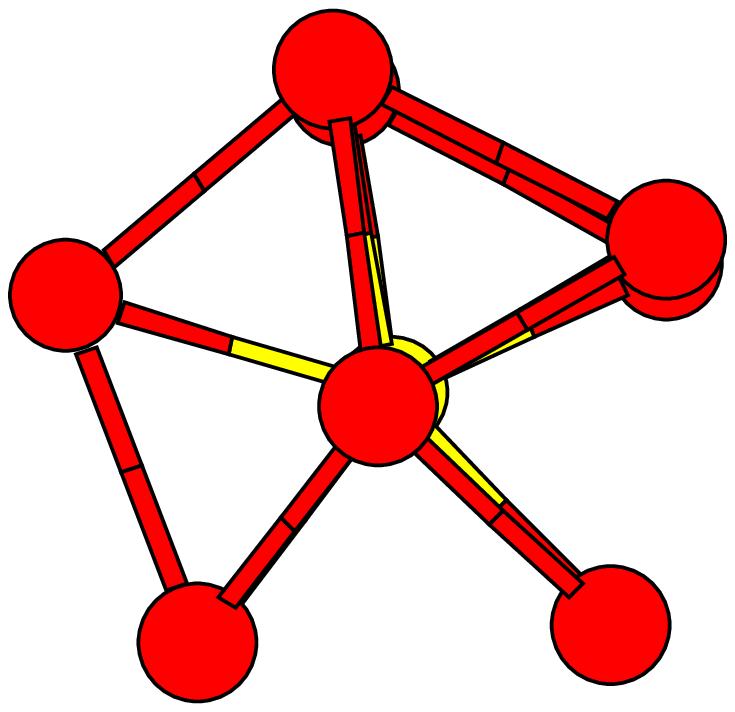}
  }
\caption{The core structures for $Ag_{13}Au_{42}$}\label{coregeomag13au42}
\end{figure}

\begin{figure}[!htb]
\centering
\subfigure[]
{
  \includegraphics[width=1.8in,height=1.40in]{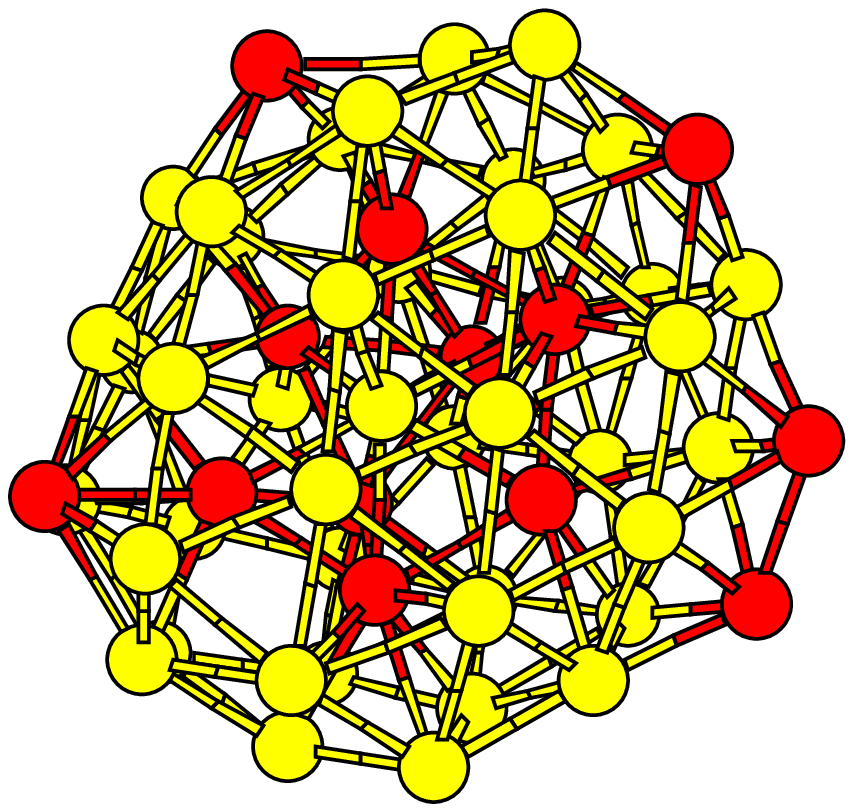}
  }
\subfigure[]
{
  \includegraphics[width=1.8in,height=1.40in]{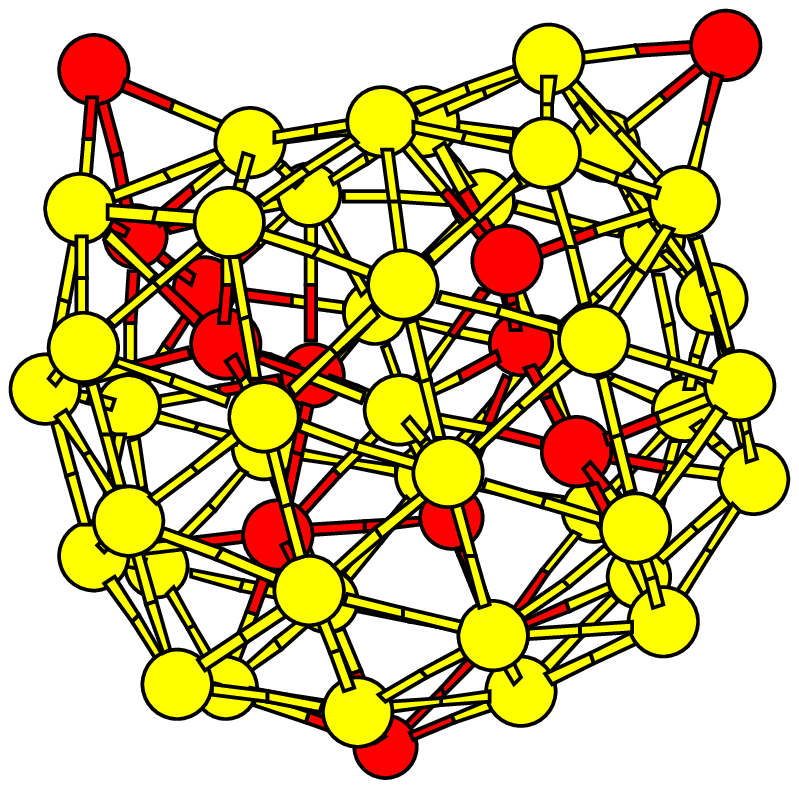}
  } \\
\subfigure[]
{
  \includegraphics[width=1.8in,height=1.40in]{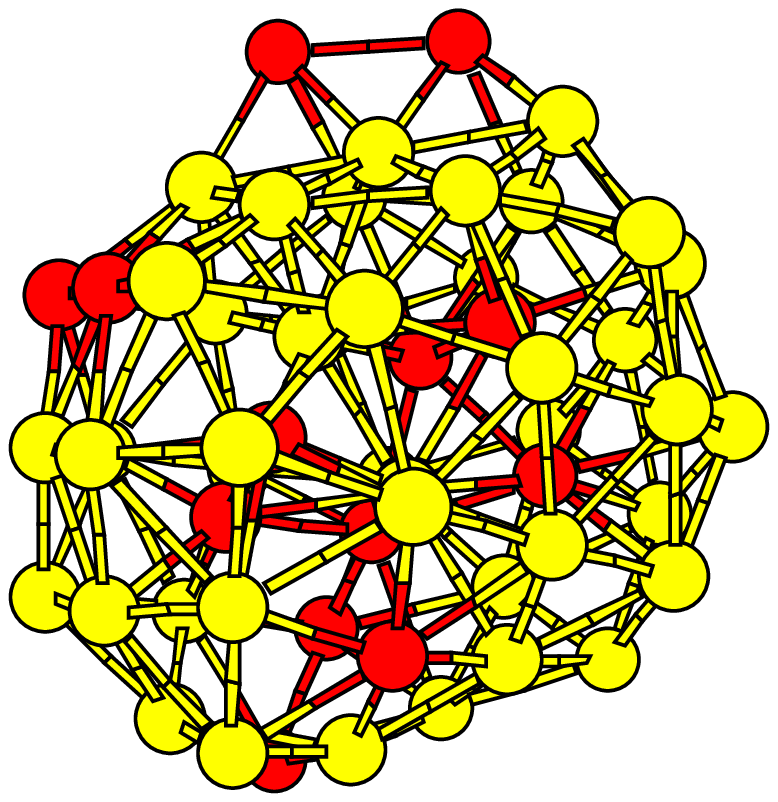}
  }
  \subfigure[]
{
  \includegraphics[width=1.8in,height=1.40in]{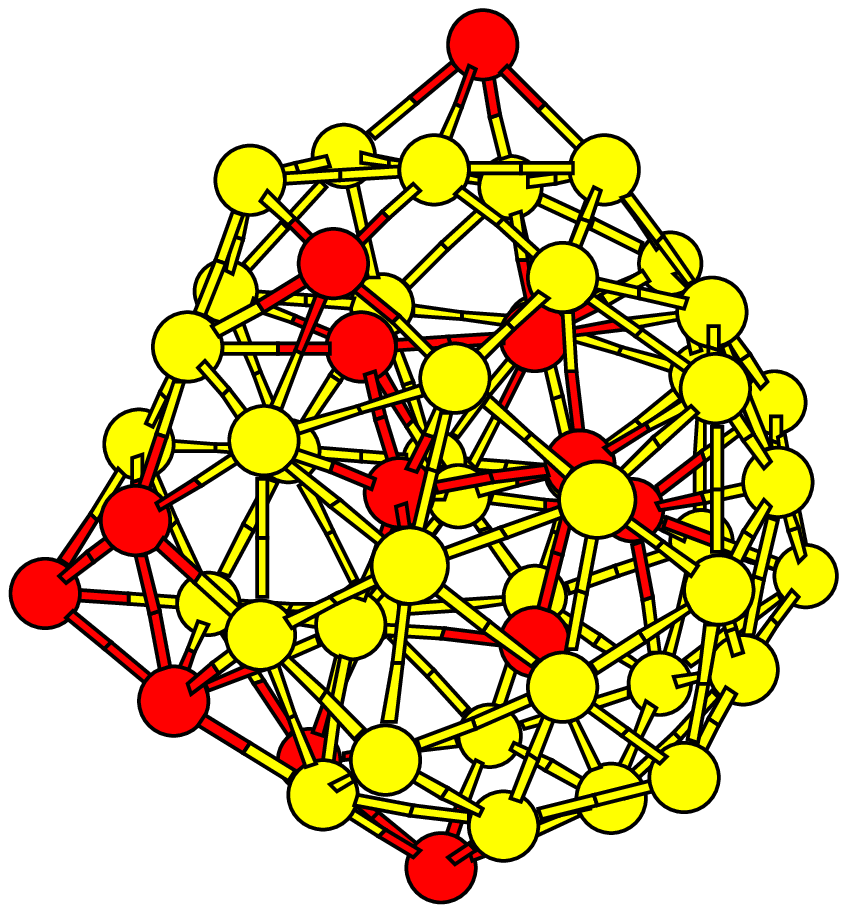}
  }
\caption{Low lying isomers of $Ag_{13}Au_{42}$}\label{someiso_sg13au42}
\end{figure}

A. $Ag_{13}Au_{42}$\\
We started BH using 10 different initial geometries. Each run was done for 30000 steps and a bunch of 50 minimum energy structures were quenched and saved from each run. A global minimum (GM) structure was obtained as shown in Fig. \ref{gm_ag13au42}. The GM structure contains an 8 atom symmetric core and 47 atom surface. It is in accordance with our previous work.\cite{agau_jcp}The 13 $Ag$ atoms are laid out as 5 on the surface and 3 in the core. We collected a total of 435 isomers which lied in an energy range of 4 eV from the obtained global minimum structure. A histogram is plotted to visualise the number of isomers obtained in an energy range as shown in Fig. \ref{histog_ag13au42}. One of the common feature among all the isomers is the presence of 5 $Ag$ atoms on the surface. Though, some high energy clusters contain more than 5 $Ag$ atoms on the surface. Due to an accurate fitting of forces, we got very different geometries of the inner core within close energy difference from the GM as shown in Fig. \ref{coregeomag13au42} and Table. \ref{tab_diffegy}. It shows the highly fluxional nature of gold-silver nanoalloys. Other than core geometries, surface atom arrangements also showed a lot of fluctuations as seen in Fig. \ref{someiso_sg13au42}. 
\begin{table}[ht]
\caption{Difference between the energies of GM structure and some low lying isomers of $Ag_{13}Au_{42}$}
\begin{center}
\begin{tabular}{|c|c|c|c|c|c|c| } 
\hline
Difference & Fig. \ref{coregeomag13au42}(a) & Fig. \ref{coregeomag13au42}(b) & Fig. \ref{coregeomag13au42}(c) \\
\hline
\multirow{2}{10em} {$\Delta$E($E_{GM}-E_{iso}$) (eV) \\}
& 0.2359 & 0.3249 & 0.3743\\
\hline
\end{tabular}
\end{center}
\label{tab_diffegy}
\end{table}\\

\begin{figure}[!htb]
\centering
\subfigure[]
{
  \includegraphics[width=1.8in,height=1.40in]{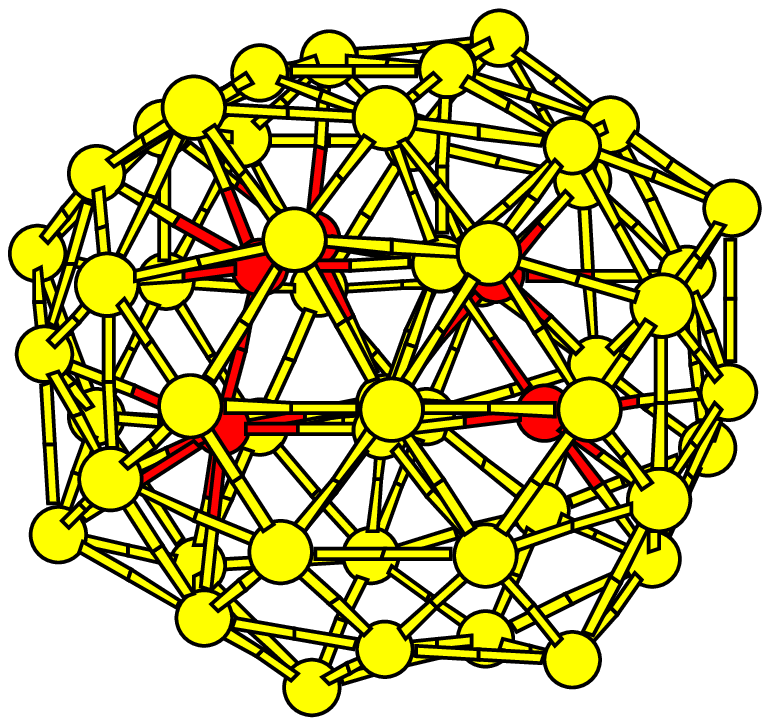}
  }
\subfigure[]
{
  \includegraphics[width=1.4in,height=1.0in]{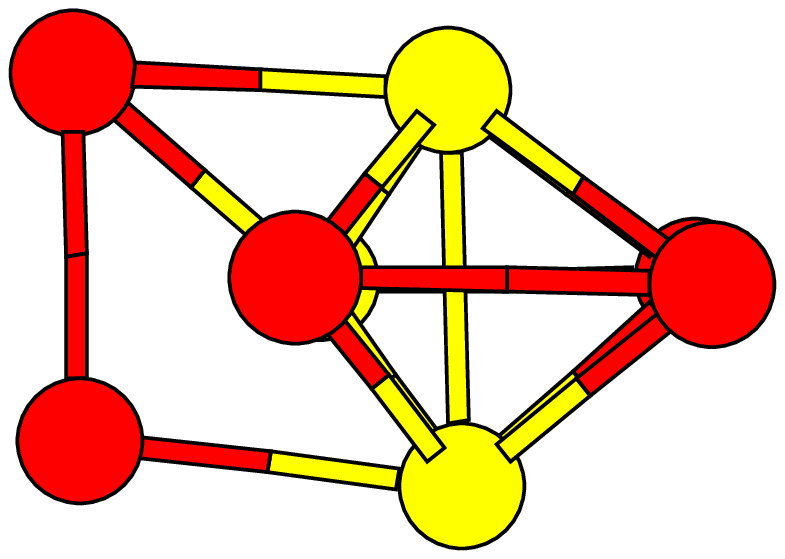}
  } 
\caption{(a) The global minimum structure of $Ag_{5}Au_{50}$, (b) The core structure of the global minimum of $Ag_{5}Au_{50}$}\label{gm_ag5au50}
\end{figure}

\begin{figure}[!htb]
\centering
\subfigure[]
{
  \includegraphics[width=1.4in,height=1.0in]{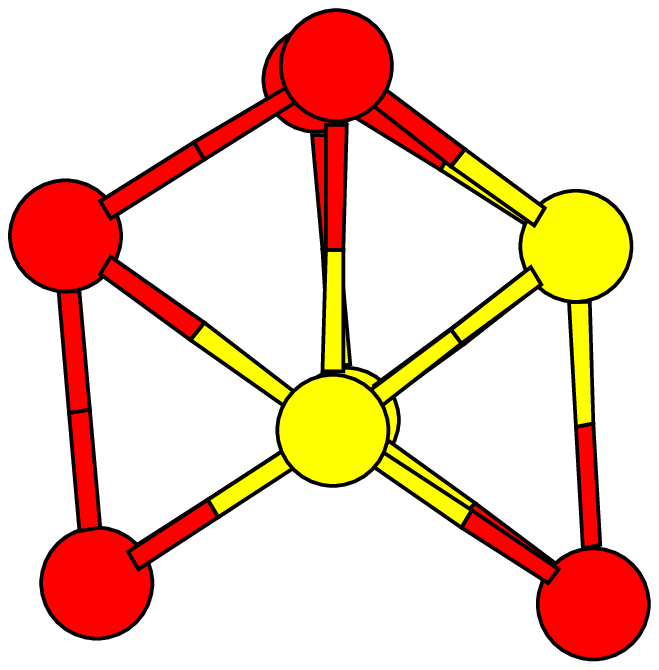}
  }
\subfigure[]
{
  \includegraphics[width=1.4in,height=1.0in]{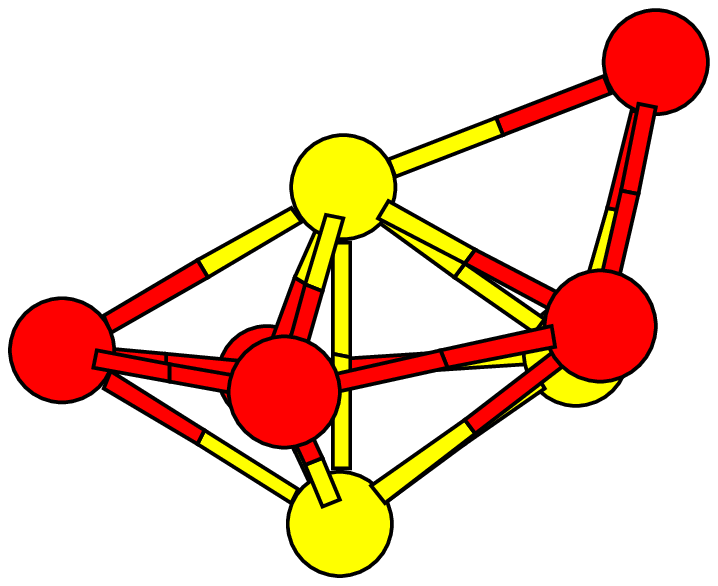}
  } 
\caption{The core atom arrangement of $Ag_{5}Au_{50}$}\label{core_ag5au50}
\end{figure}

\begin{figure}[!htb]
\centering
\subfigure[]
{
  \includegraphics[width=1.8in,height=1.40in]{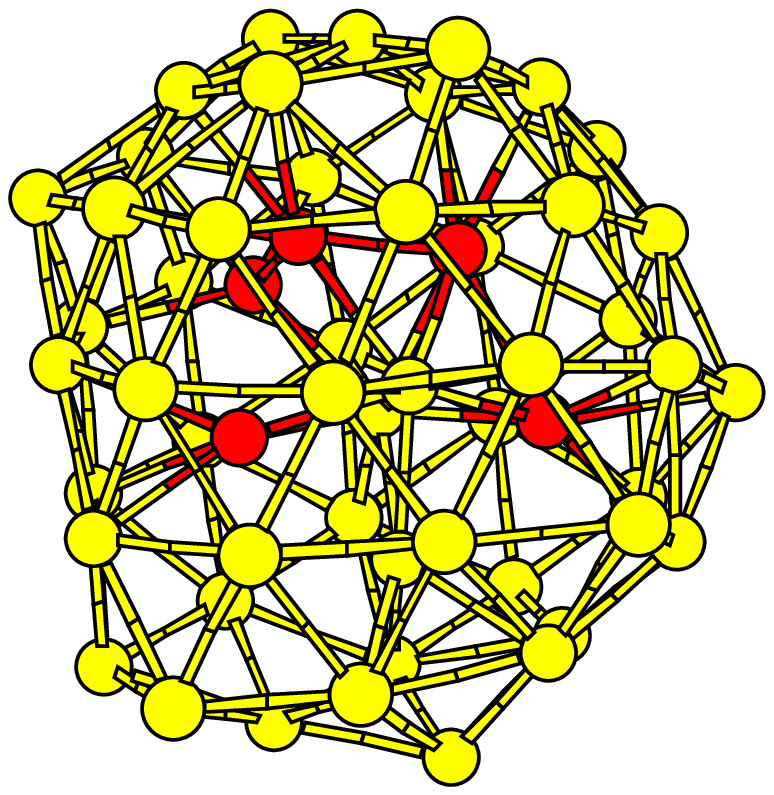}
  }
\subfigure[]
{
  \includegraphics[width=1.8in,height=1.40in]{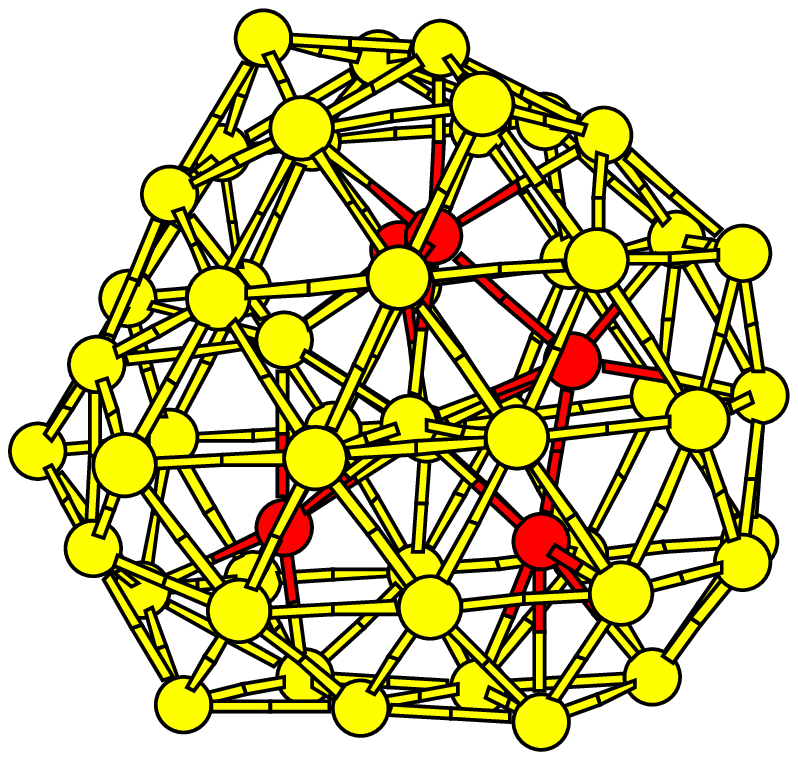}
  } \\
\subfigure[]
{
  \includegraphics[width=1.8in,height=1.40in]{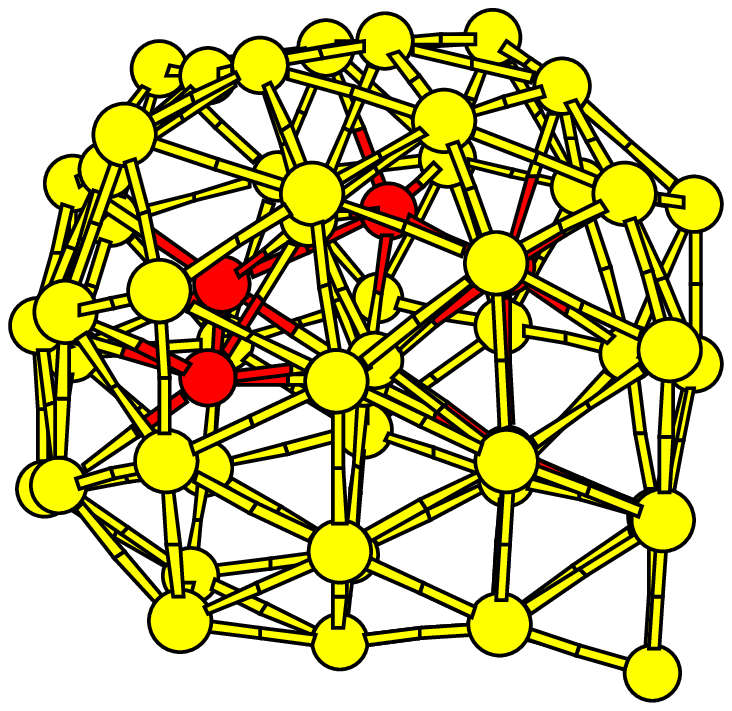}
  }
  \subfigure[]
{
  \includegraphics[width=1.8in,height=1.40in]{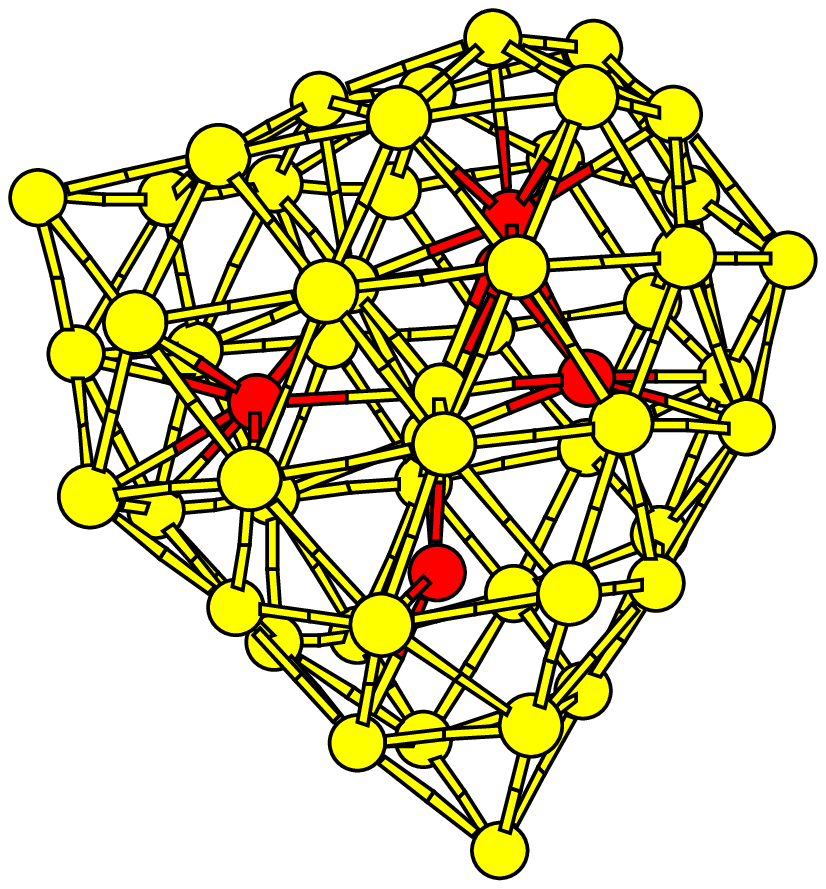}
  }
\caption{Low lying isomers of $Ag_{5}Au_{50}$ $\Delta$E($E_{GM}-E_{iso}$) (a) 0.1212 eV, (b) 0.3407 eV, (c) 0.5259 eV and (d) 1.16 eV}\label{someiso_ag5au50}
\end{figure}

B. $Ag_{5}Au_{50}$\\
For exploring the PES of $Ag_{5}Au_{50}$, we ran MDS at 300 K and 400 K for a total time of 1 $ns$ at a time step of 2 $fs$ using different initial structures. We quenched the structure after ever 20 $ps$ of the simulation. We got a GM isomer having an amorphous surface as shown in Fig. \ref{gm_ag5au50}(a). A symmetric core arrangement was observed in the GM structure as shown in Fig. \ref{gm_ag5au50}(b). Since, Au atoms are in majority, the structure is more towards amorphous. We got a lot of low lying isomers within a narrow energy range of 1 $eV$ from the found GM structure. Different core atoms arrangement was discovered in the low lying isomers as shown in Fig. \ref{core_ag5au50}. The difference between the energy of isomers shown in Fig. \ref{core_ag5au50}(a) and (b) from the GM is 0.0118 eV and 0.4856 eV, respectively. Since, the energy difference is less than 0.5 eV, it shows that gold doped nanoclusters are fluxional in nature. A lot of isomers were identified with very different arrangement of surface atoms as shown in Fig. \ref{someiso_ag5au50}.

\begin{figure}[!htb]
\centering
\subfigure[]
{
  \includegraphics[width=1.8in,height=1.40in]{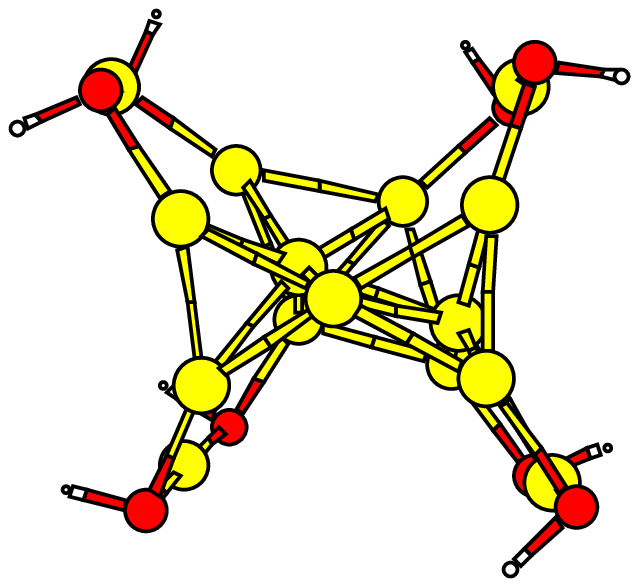}
  }
\subfigure[]
{
  \includegraphics[width=1.8in,height=1.40in]{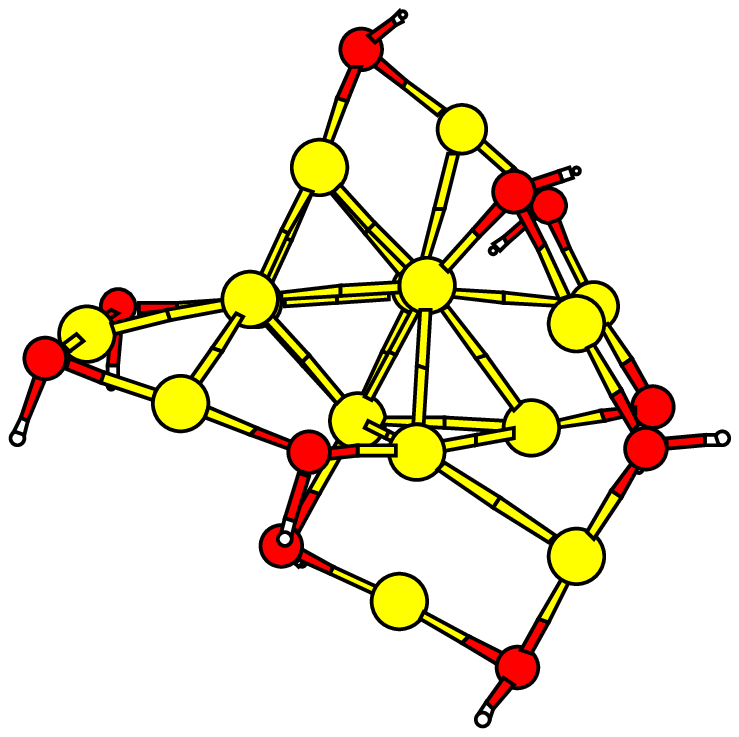}
  } \\
\subfigure[]
{
  \includegraphics[width=1.8in,height=1.40in]{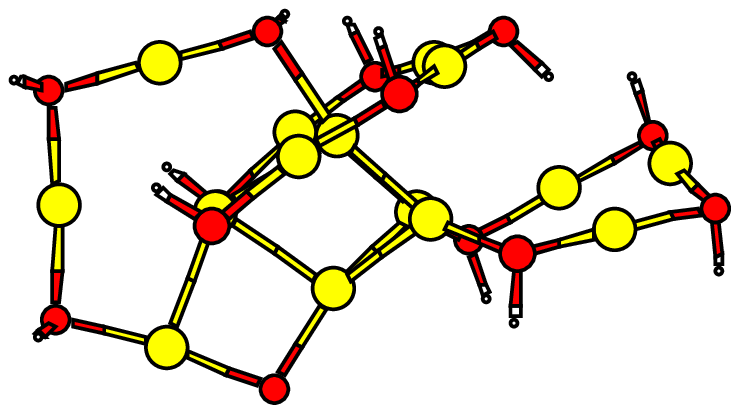}
  }
\caption{GM structures of (a) $Au_{15}(SH)_{8}$, (b) $Au_{15}(SH)_{10}$, (c) $Au_{15}(SH)_{13}$,}\label{gmiso_thio}
\end{figure}

C. $Au_{15}(SH)_{8}$, $Au_{15}(SH)_{10}$ and $Au_{15}(SH)_{13}$\\
We selected thio- protected $Au_{15}$ cluster with different concentration of $SH$ group to study the GO. We ran MDS at 100 K and 150 K for sampling the PES of thio- protected gold clusters. We sampled 212 structures for $Au_{15}(SH)_8$ and 208 structures for $Au_{15}(SH)_{10}$ in an energy interval of 0.5 eV from the tentative GM structure obtained from the MDS. For $Au_{15}(SH)_{13}$, we sampled 220 structures in an energy interval of 1.0 eV from the tentative GM structure. All the simulations were run at time step of 1 $fs$ and the total time of simulation was 2 $ns$. The GM structures are shown in Fig. \ref{gmiso_thio}. We observed that as the number of units of $SH$ increased from 8 to 10, a more symmetric structure is obtained. But, as we increased the units to 13, there was not much impact on the symmetry of the structure.

\begin{figure}[!htb]
\centering
\subfigure[]
{
  \includegraphics[width=1.8in,height=1.40in]{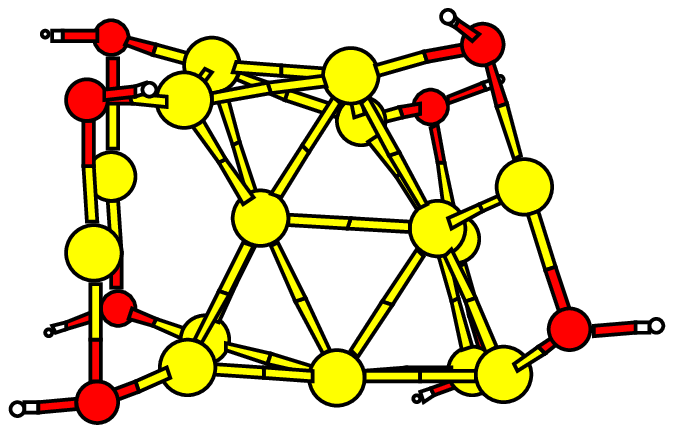}
  }
\subfigure[]
{
  \includegraphics[width=1.8in,height=1.40in]{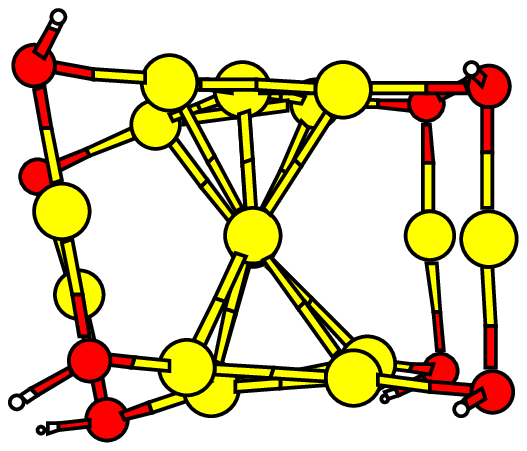}
  } \\
\subfigure[]
{
  \includegraphics[width=1.8in,height=1.40in]{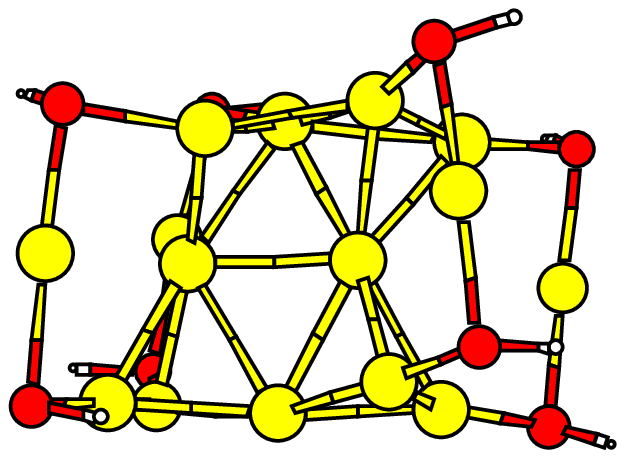}
  }
\caption{Low lying isomers of $Au_{15}(SH)_{8}$, $\Delta$E ($E_{GM}-E_{iso}$) (a) 0.2389 eV, (b) 0.2737 eV, (c) 0.4178 eV}\label{lowiso_sh8}
\end{figure}

\begin{figure}[!htb]
\centering
\subfigure[]
{
  \includegraphics[width=1.8in,height=1.40in]{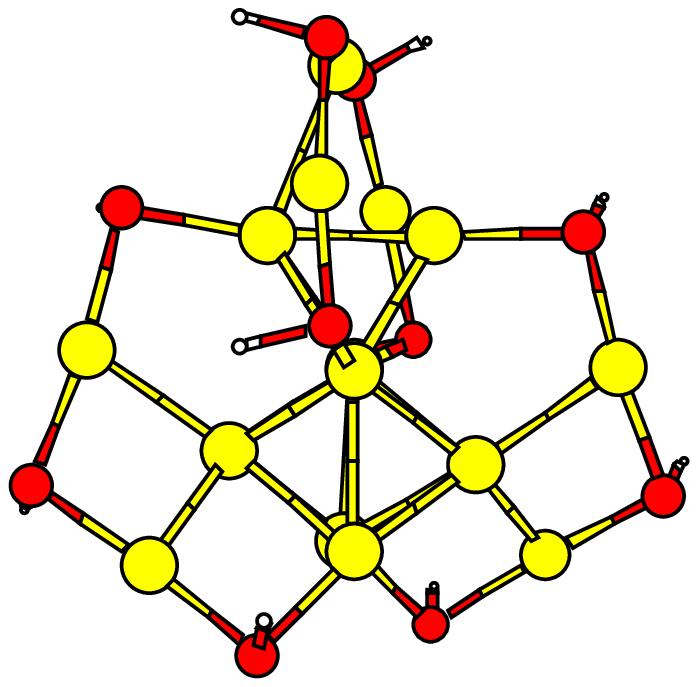}
  }
\subfigure[]
{
  \includegraphics[width=1.8in,height=1.40in]{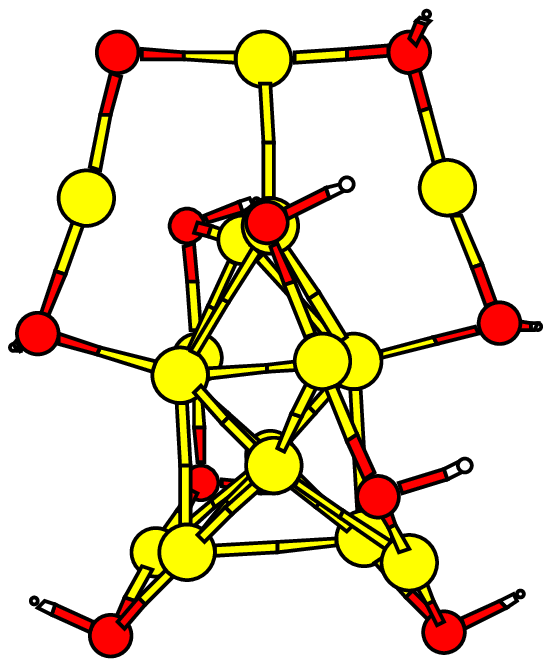}
  } \\
\subfigure[]
{
  \includegraphics[width=1.8in,height=1.40in]{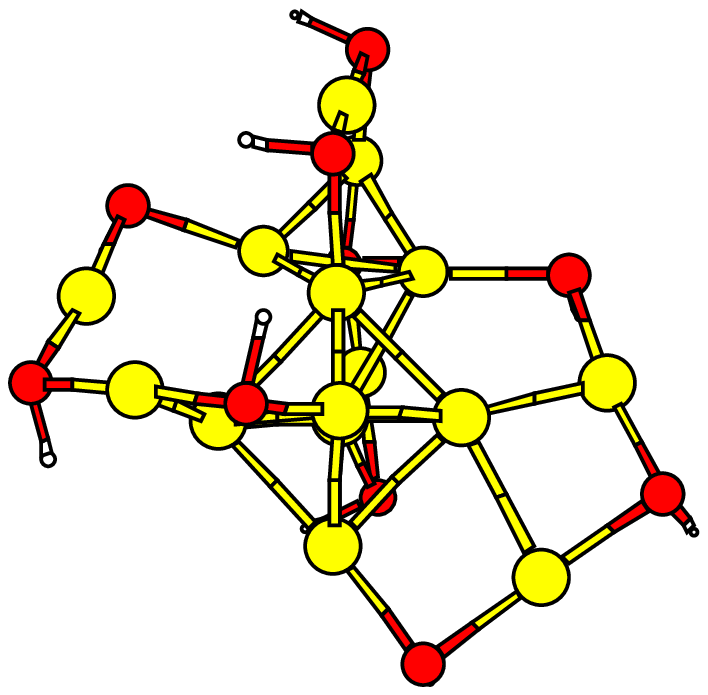}
  }
\caption{Low lying isomers of $Au_{15}(SH)_{10}$, $\Delta$E ($E_{GM}-E_{iso}$) (a) 0.005 eV, (b) 0.095 eV, (c) 0.205 eV}\label{lowiso_sh10}
\end{figure}

\begin{figure}[!htb]
\centering
\subfigure[]
{
  \includegraphics[width=1.8in,height=1.40in]{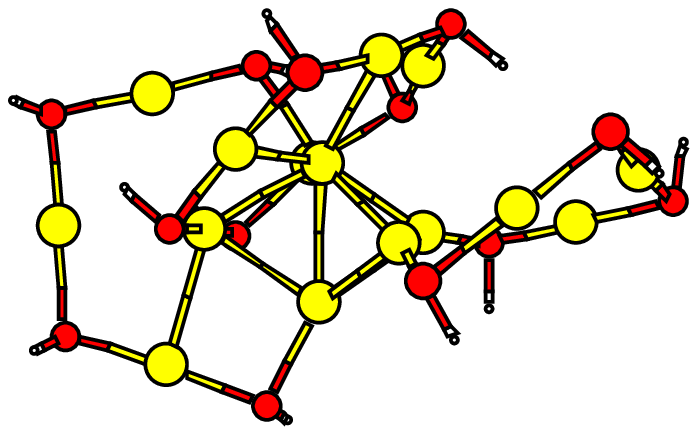}
  }
\subfigure[]
{
  \includegraphics[width=1.8in,height=1.40in]{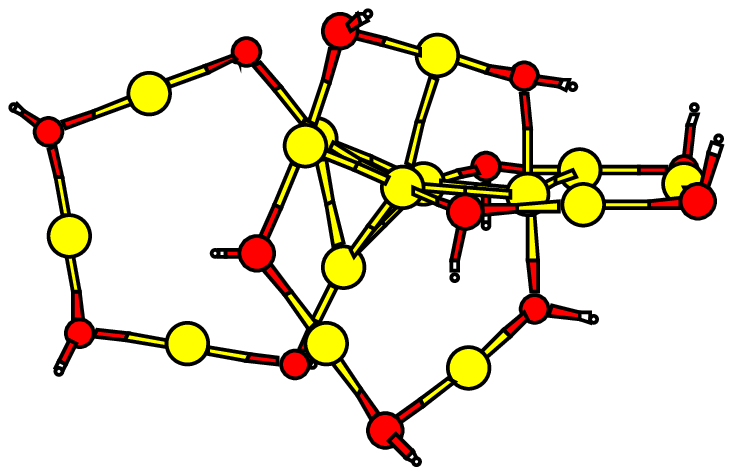}
  } 
\caption{Low lying isomers of $Au_{15}(SH)_{13}$, $\Delta$E ($E_{GM}-E_{iso}$) (a) 0.614 eV, (b) 1.329 eV}\label{lowiso_sh13}
\end{figure}

We obtained different geometry isomers for all the three compositions considered. The low lying isomers for $Au_{15}(SH)_8$, $Au_{15}(SH)_{10}$ and $Au_{15}(SH)_{13}$ are shown in Fig. \ref{lowiso_sh8}, \ref{lowiso_sh10}, \ref{lowiso_sh13}, respectively. A conclusion that can be made from the different isomers obtained for silver-gold nanoalloys and thio- protected gol nanoclusters is that gold based nanoparticles exhibit a lot of fluctuations in their structure and thus reactivities can be tuned according to different geometries obtained. 
\section{Conclusions}\label{conclu}
We have proposed a transferable ANN model for fitting energy and forces for any nanoparticle. Our proposed strategy is termed as transferable due to the model dependence on inter-atomic distances, effective nuclear charges of the chemical species and reduced mass of the bonds involved in a chemical system. By doing a concurrent but decoupled fitting of energy and forces of a MC system using a SANN, we get an accurate representation of the atomic environments. The usage of same network for any chemical species in the system leads to a tremendous reduction in the computational costs. Since, forces are highly sensitive for an atomic environment, its fitting was a challenge, which was addressed in this work. We applied our proposed method to two systems. One consisting of a bimetallic alloy i.e. $(AgAu)_{55}$-$(AgAu)_{147}$ and the other a range of thiol protected gold nanoclusters ($Au_{13}(SH)_{6}$-$Au_{38}(SH)_{24}$), $Au_{68}(SH)_{32}$. Alike $Au_{147}$,\cite{sphharmjcp} the icosahedron geometry is not favorable as a local minimum structure for $Ag_{35}Au_{112}$. Due to a precise representation of forces, the weights obtained for the dataset consisting of $Au_{13}(SH)_{6}$-$Au_{38}(SH)_{24}$ were able to optimize the geometry of $Au_{68}(SH)_{32}$ and run its dynamics. Our proposed method can help in studying the global optimizations and dynamics of many other MC systems consisting of nanoparticles.
\begin{acknowledgements}
We thank IIT Indore for providing server facilities - OMICORN [GenuineIntel 2600.0 MHz], GOLD [AMD Opteron(TM) 2600.0 MHz]. SJ thanks IIT Indore for research fellowship.
\end{acknowledgements}
\bibliographystyle{phaip}

\bibliography{multicom}

\end{document}